\documentclass[lettersize,journal]{IEEEtran}
\usepackage{amsmath,amsfonts}
\usepackage{algorithmic}
\usepackage{algorithm}
\usepackage{array}
\usepackage[caption=false,font=normalsize,labelfont=sf,textfont=sf]{subfig}
\usepackage{textcomp}
\usepackage{stfloats}
\usepackage{url}
\usepackage{verbatim}
\usepackage{graphicx}
\usepackage{cite}
\usepackage[table]{xcolor}  % 
\usepackage{colortbl}        % 
\usepackage{multirow}
\usepackage{booktabs}
\begin{document}

\title{Generating Human-AI Collaborative Design Sequence for 3D Assets via Differentiable Operation Graph}

\author{Xiaoyang Huang, Bingbing Ni, Wenjun Zhang
        % <-this % stops a space
\thanks{Corresponding author: Bingbing Ni.}% <-this % stops a space
\thanks{ The authors are with the Shanghai Jiao Tong University, Shanghai 200240, China (e-mail: huangxiaoyang@sjtu.edu.cn; nibingbing@sjtu.edu.cn; zhangwenjun@sjtu.edu.cn;}
}

% The paper headers
% \markboth{Journal of \LaTeX\ Class Files,~Vol.~14, No.~8, August~2021}%
% {Shell \MakeLowercase{\textit{et al.}}: A Sample Article Using IEEEtran.cls for IEEE Journals}

% \IEEEpubid{0000--0000/00\$00.00~\copyright~2021 IEEE}
% Remember, if you use this you must call \IEEEpubidadjcol in the second
% column for its text to clear the IEEEpubid mark.

\maketitle

\begin{abstract}
The emergence of 3D artificial intelligence-generated content (3D-AIGC) has enabled rapid synthesis of intricate geometries. However, a fundamental disconnect persists between AI-generated content and human-centric design paradigms, rooted in representational incompatibilities: conventional AI frameworks predominantly manipulate meshes or neural representations (\emph{e.g.}, NeRF, Gaussian Splatting), while designers operate within parametric modeling tools. This disconnection diminishes the practical value of AI for 3D industry, undermining the efficiency of human-AI collaboration. To resolve this disparity, we focus on generating design operation sequences, which are structured modeling histories that comprehensively capture the step-by-step construction process of 3D assets and align with designers' typical workflows in modern 3D software. We first reformulate fundamental modeling operations (\emph{e.g.}, \emph{Extrude}, \emph{Boolean}) into differentiable units, enabling joint optimization of continuous (\emph{e.g.}, \emph{Extrude} height) and discrete  (\emph{e.g.}, \emph{Boolean} type) parameters via gradient-based learning. Based on these differentiable operations, a hierarchical graph with gating mechanism is constructed and optimized end-to-end by minimizing Chamfer Distance to target geometries. Multi-stage sequence length constraint and domain rule penalties enable unsupervised learning of compact design sequences without ground-truth sequence supervision. Extensive validation demonstrates that the generated operation sequences achieve high geometric fidelity, smooth mesh wiring, rational step composition and flexible editing capacity, with full compatibility within design industry. 
\end{abstract}


% According to a comprehensive user study with 407 professional designers, 90\% of participants reported over 60\% time savings compared to fully manual modeling when using our method, with 70\% of participants indicating over 80\% time savings. In contrast, more than 72\% experienced no time savings or even increased time costs using current 3D-AIGC methods in 3D projects. 

\section{Introduction}\label{sec1}

Recent advances in 3D-AIGC systems have achieved remarkable progress in generating intricate 3D assets from multimodal inputs, demonstrating unprecedented geometric complexity and visual fidelity \cite{lee2024dreamflow, babu2024hyperfields}. Despite these technical breakthroughs, a fundamental disconnect persists between AI-generated content and professional 3D production workflows, particularly regarding parametric control and procedural editing capabilities \cite{jones2023shapecoder, jones2021shapemod, jones2020shapeassembly}. This incompatibility originates from fundamental representational disparities: professional artists typically construct assets through hierarchical procedural operations in Digital Content Creation (DCC) platforms like Blender \cite{raistrick2023infinite, raistrick2024infinigen, huang2024blenderalchemy, gu2025blendergym}, while contemporary neural methods predominantly optimize surface representations using meshes \cite{chen2024meshanything,chen2024meshanythingv2,siddiqui2024meshgpt}, NeRF \cite{mildenhall2021nerf, pooledreamfusion}, or Gaussian Splatting \cite{kerbl20233d, yi2023gaussiandreamer}. The practical implications become apparent when modifying AI-generated content, where the lack of construction history forces artists into labor-intensive vertex manipulation rather than adjusting original design parameters, akin to repainting dried oil versus revising digital sketch layers in 2D art production. Our user study with professional 3D designers reveals significant inefficiencies in adapting 3D-AIGC assets: 70\% of practitioners spent over 3 hours modifying 3D-AIGC results for production integration. Notably, more than 72\% reported no time savings or even increased time costs compared to manual modeling from scratch.

% \IEEEpubidadjcol

To bridge this gap, we unify 3D asset representations with designers' typical workflow by adopting design operation sequences, the structured records of the procedural history through which 3D assets are constructed. These sequences encapsulate the complete evolution of a digital asset as timestamped operations with associated parameters (\emph{e.g.}, [\emph{AddCube} → \emph{Subdivision} → \emph{Extrude} → \emph{Bevel}], with parameters abstracted). This approach mirrors modern DCC workflows, preserving both geometric transformations and design rationale through all developmental stages. Building upon this foundational alignment, we formulate the problem as the decomposition and reconstruction of these design operation sequences. While prior research has predominantly focused on CAD command sequences \cite{khan2024cad, ren2022extrudenet, xu2022skexgen, liu2024split, li2022free2cad, seff2021vitruvion} and analytical primitives \cite{yu2022capri, yavartanoo20213dias, huang2023learning, chen2020bsp, chen2019bae, liu2022robust, deng2020cvxnet, paschalidou2020learning, paschalidou2019superquadrics, kawana2020neural, paschalidou2021neural} to represent 3D objects, these approaches suffer from two critical limitations: constrained applicability to diverse 3D assets and incompatibility with standard DCC workflows. In contrast, design operation sequences offer significantly broader generalization capabilities while maintaining strong alignment with professional designers' established workflow practices.

Modeling 3D assets by operation sequence has evident advantages in human-AI collaborative production pipelines. First, artists could adjust foundational parameters in an AI-generated asset's operation history rather than manipulating dense polygon soups. This operational decomposition becomes particularly valuable when assets require late-stage modifications, as procedural models enable step-independent adjustments without global re-engineering. In animation pipelines requiring iterative character variants, accessible modeling sequences facilitate efficient modifications. For example, a creature designer could refine an AI-generated dragon model by tweaking specific \emph{Extrude} parameters in its operation history to adjust wing proportions, rather than manually sculpting mesh deformations.
Second, the mesh wiring regularity in operation-based modeling is crucial in production environments. Models built through procedural operations inherently maintain cleaner topology \cite{yang2023topology} and structured edge flow \cite{takayama2013sketch, marcias2015data} compared to neural implicit methods. This geometric discipline proves essential for multiple downstream workflows. For example, physics simulations like cloth dynamics \cite{yu2024super} require evenly distributed mesh elements for accurate collision resolution; deformation in character articulation requires edge loops to follow biomechanical structures for volume preservation during animation \cite{nuvoli2022skinmixer}, while neural-reconstructed character models usually fail animation tests due to topological inconsistencies. Even for static asset creation, structured edge flow enables efficient detail sculpting \cite{de2017regularized} and texture baking \cite{knodt2023joint} through logical UV space partitioning.

\begin{figure}[!t]
    \includegraphics[width=\linewidth]{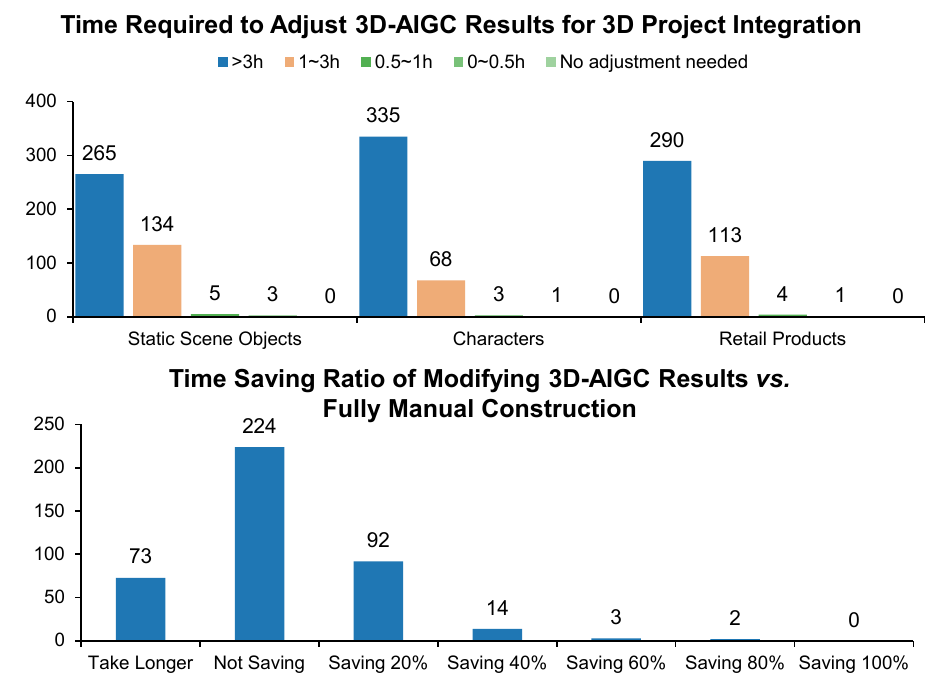}
    \caption{Our user study with professional 3D designers reveals that 70\% of practitioners spent over 3 hours modifying 3D-AIGC outputs for production integration, while more than 72\% reported no time savings or even increased time costs compared to manual modeling from scratch.
    }
    \label{fig:survey_intro}
\end{figure}

The technical challenges in reconstructing modeling sequences originate from two principal factors. The first challenge arises from the extreme heterogeneity of modeling operations. DCC software like Blender provide dozens of fundamental operations (\emph{e.g.}, \emph{Extrude}, \emph{Bevel}, \emph{Subdivision}) that involve both continuous parameter adjustments (\emph{e.g.}, bevel width from 0.1mm to 10mm) and discrete selections (\emph{e.g.}, \emph{Boolean} operation modes ranging in union, difference and intersection). This hybrid parameter space exhibits non-convex optimization landscapes with discontinuous boundaries. Beside, its combinatorial complexity scales exponentially with sequence length. For example, a modeling sequence with L operations yields approximately $10^{2L}$ parameter combinations (assuming 2 key parameters per operation). In character modeling, a basic animal head model might sequentially undergo \emph{Mirror} modifiers, surface subdivision, face inset, and extrusion, where both operational ordering and parametric combinations critically determine morphological outcomes.

The second challenge manifests in the conflict between data scarcity and algorithmic generalization requirements. While the 3D modeling community maintains extensive public repositories of 3D asset resources, complete procedural histories (\emph{e.g.}, .blend project files) remain exceptionally scarce. Many commercial models involve proprietary construction processes that rarely enter the public domain. Moreover, the acquisition of operation sequence data itself represents a computationally expensive and labor-intensive endeavor, requiring exhaustive documentation of every operational step executed by designers during the creative workflow. Existing large-scale operation sequence datasets (\emph{e.g.}, ABC dataset\cite{koch2019abc}, DeepCAD \cite{wu2021deepcad}, CAD-MLLM \cite{xu2024cad}) are strictly limited to mechanical components, covering only a narrow range of operations (\emph{e.g.}, \emph{Extrude}) and asset categories, which exhibit limited generalizability to broader modeling tasks. Under these prevailing circumstances, there appear to be no viable prospects for acquiring a sufficient quantity of operation sequence data required to effectively train data-intensive models, such as transformer-based architectures. Such limitations have compelled us to develop an operation sequence modeling algorithm that does not require large-scale training data.

To this end, we propose a zero-shot approach which extracts operation sequences from arbitrary 3D assets using a differentiable operation framework. Our method unifies  procedural modeling with neural optimization by reformulating conventional geometric operations into differentiable components, enabling joint optimization of both continuous and discrete parameters through gradient-based learning. 
The framework organizes operations into a hierarchical computational graph, where each node represents a geometric transformation (\emph{e.g.}, extrusion, rotation) parameterized by continuous variables (\emph{e.g.}, angles, displacements) and discrete choices (\emph{e.g.}, operation types, segment counts). Since discrete choices could not be optimized directly via chain rule like continuous parameters, we address them through a probabilistic branching mechanism, which performs parallel evaluation of candidate operations followed by weighted expectation calculation, allowing gradient propagation during training and deterministic selection during inference. By associating nodes with learnable activation coefficients, a gating mechanism dynamically modulates the influence of each operation node during optimization. 
The graph is trained end-to-end to minimize geometric discrepancies between generated and target shapes, quantified by Chamfer Distance of sampled point clouds. This unified optimization simultaneously updates continuous parameters, discrete operation weights, and node activation. 
At inference, the system automatically prunes redundant operations whose activation falls below a threshold. The final operation sequence, along with its optimized parameters, is extracted by retaining activated nodes and resolving discrete choices to their highest-probability configurations. 
This approach preserves geometric precision while maintaining procedural editability. Furthermore, the zero-shot paradigm eliminates the need for pre-training on large-scale labeled operation sequences.

\begin{figure*}[!t]
    \includegraphics[width=\linewidth]{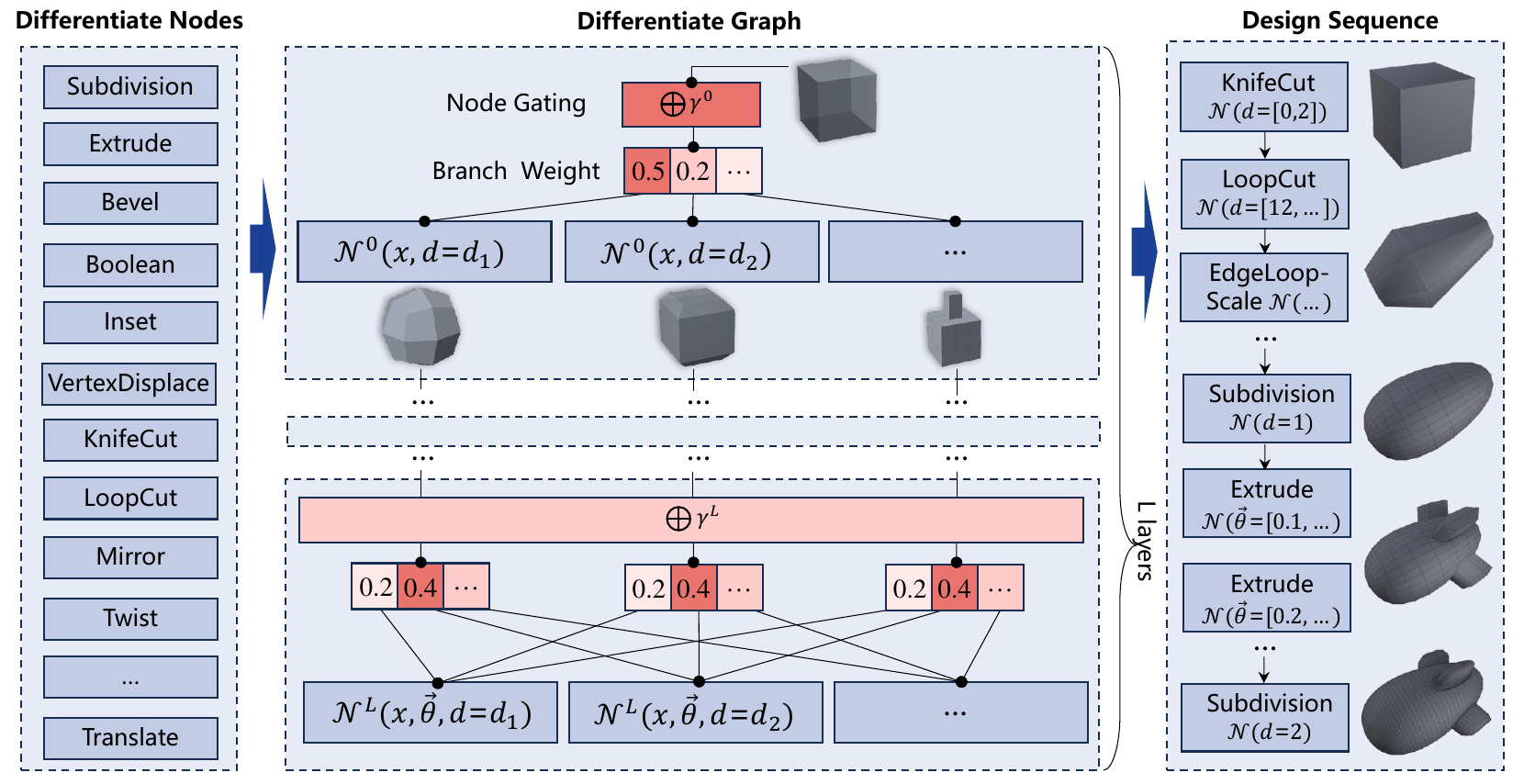}
    \caption{Illustration of our framework: First, conventional 3D modeling operations are systematically reformulated into differentiable nodes, forming the foundation for gradient-based optimization. Next, a hierarchical, differentiable operation graph is constructed by stacking multiple operation nodes controlled by gating and probabilistic branching mechanisms. Finally, optimized design operation sequences are extracted from the graph. This framework thereby bridges neural optimization techniques with procedural modeling, enabling seamless integration with human-centric 3D modeling workflows.
    }
    \label{fig:system}
% \vspace{-0.15in}
\end{figure*}

Notably, all operations in the developed framework directly correspond to native modeling tools in popular 3D design software, such as Blender, 3dsMax and Maya. This advancement fundamentally redefines human-AI collaboration in 3D content creation. By bridging neural generation techniques with procedural modeling paradigms, artists gain the generative capabilities of AI while preserving the controllability of manual workflows. As AI-assisted workflows proliferate in digital content creation, our approach addresses the critical challenge of maintaining artistic authority while harnessing algorithmic efficiency through human-aligned procedural representations.

% To evaluate our algorithm’s impact on human-AI collaboration in 3D design, we conducted a comprehensive questionnaire-based user study with 407 qualified 3D modeling professionals recruited via industry forums, networks and academic collaborations. The survey instrument is composed of five interconnected modules with 31 closed and open-ended items. Module 1 captured participant background for segmentation. Module 2 identified pain points in current tools (e.g., fidelity, mesh topology, editability) and quantified time savings estimates across object types (static objects, characters, products). Module 3 performed a comparative analysis against key baselines across seven criteria including fidelity, mesh wiring, step rationality, and workflow compatibility, revealing automation-usability trade-offs. Module 4 specifically rated the proposed method's compliance with core requirements and time efficiency. Module 5 gathered subjective feedback on procedural decomposition impact, adoption willingness, and workflow enhancement potential. Employing cognitive pilot testing and expert validation ensured methodological robustness.

Our method enables zero-shot extraction of design operation sequences from arbitrary 3D assets. Experimental validation shows that the derived operation sequences achieve precise geometric reconstruction and high-quality mesh wiring. Superior human-AI synergy is demonstrated through step-by-step decomposition aligning with the ground-truth modeling sequence and workflow-compatible parametric editing.

\section{Related Work}

\subsection{3D-AIGC}

Recent advances in 3D-AIGC systems have demonstrated unprecedented capabilities in generating intricate 3D assets from multimodal inputs, achieving breakthroughs in geometric complexity and photorealistic fidelity \cite{lee2024dreamflow, babu2024hyperfields}. Generative adversarial networks (GANs) have long served as foundational frameworks for 3D synthesis, with innovations such as GET3D \cite{gao2022get3d} integrating differentiable rendering with 2D GANs to generate textured meshes, EG3D \cite{chan2022efficient} adopting triplane representations for enhanced 3D-GAN efficiency, and NeRF-based approaches including GRAF \cite{schwarz2020graf} and GIRAFFE \cite{fu2022representing} enabling novel-view synthesis. Concurrently, diffusion-based methods have accelerated substantially: DreamFusion \cite{poole2022dreamfusion} bridges NeRF with 2D diffusion for text-to-3D generation, Point-E \cite{nichol2022point} employs cascaded diffusion for efficient point cloud synthesis, Shap·E \cite{shue20233d} leverages triplane-encoded data for 2D diffusion training, and Rodin \cite{wang2023rodin} projects NeRF features into 2D planes for 3D-aware diffusion.

Despite these advancements, contemporary neural methodologies predominantly optimize surface representations through meshes \cite{chen2024meshanything,chen2024meshanythingv2,siddiqui2024meshgpt}, NeRF \cite{mildenhall2021nerf, pooledreamfusion}, or Gaussian Splatting \cite{kerbl20233d, yi2023gaussiandreamer}. While existing 3D-AIGC systems \cite{zhao2025hunyuan3d, xiang2024structured, hyper3d_ai} deliver high-precision photorealistic outputs, they exhibit critical limitations in design-industry applications. Their outputs are constrained to final-state meshes that lack parametric flexibility for procedural modification aligned with designers' creative intents, fundamentally mismatching professional workflows. Furthermore, these assets cannot integrate with standard modeling/animation pipelines without labor-intensive topological rebuilding, remaining largely restricted to rendering purposes rather than functioning as editable components.

To address these gaps, our work establishes an industry-compatible framework for intelligent 3D content creation that natively supports professional design workflows. By transcending the limitations of static mesh outputs, our approach enables procedural adaptability and seamless integration with production pipelines.

% \subsection{Sequential Modeling}

% There have also been some explorations on formulating 3D assets creation as a sequential operation modeling problem in computer graphics. Among them, visual program generation is deeply investigated in ShapeAssembly~\cite{jones2020shapeassembly} and its successors, ShapeMod~\cite{jones2021shapemod} and ShapeCoder~\cite{jones2023shapecoder}, by representing a shape dataset with geometry abstraction functions and programs.

\subsection{CAD modeling}

\begin{table*}[!ht]
\centering
% \rowcolors{2}{blue!10}{blue!5}
% \setlength\tabcolsep{1pt}
% \begin{tabular*}{\hsize}{@{}@{\extracolsep{\fill}}lcccc@{}}
\caption{Design operations addressed in this work are distinct from CAD commands in terms of application scenarios, operation diversity and data availability.}
\begin{tabular}{lcc}
\toprule
% \hline

 & Design Operations & CAD commands \\
 \midrule
Application & 3D Content Creation & Computer-Aided Design \\
Target Assets & Scene Objects, Characters, Retail Products & Mechanical Components \\
Applied Industry & Films, Gaming, VR/AR, Advertising & Machinery Manufacturing \\
Platform & Blender, Maya, \emph{etc.}& AutoCAD, SolidWorks, \emph{etc.} \\
Operation Diversity & Dozens of Types &  Mainly Sketch and Extrusion \\
Data Availability & None & Millions\\
Datasets & None & ABC, DeepCAD, \emph{etc.} \\
Prior Studies & None & DeepCAD, PointCAD, \emph{etc.} \\
 
\bottomrule
% \hline
\end{tabular}
% \vspace{5pt}

\label{tab:compare_cad}
\end{table*}

% \begin{table}[!t]
% \renewcommand{\arraystretch}{1.5} 
% \centering
% % \rowcolors{2}{blue!10}{blue!5}
% \begin{tabular}{lcc}
% \toprule
% % \rowcolor{blue!20}
%  & Design Operations & CAD commands \\
% \midrule
% Application & 3D Content Creation & Computer-Aided Design \\
% Target Assets & Scene Objects, Characters, Retail Products & Mechanical Components \\
% Applied Industry & Films, Gaming, VR/AR, Advertising & Machinery Manufacturing \\
% Platform & Blender, Maya, \emph{etc.}& AutoCAD, SolidWorks, \emph{etc.} \\
% Operation Diversity & Dozens of Types &  Mainly Sketch and Extrusion \\
% Data Availability & None & Millions\\
% Datasets & None & ABC, DeepCAD, \emph{etc.} \\
% Prior Studies & None & DeepCAD, PointCAD, \emph{etc.} \\
 
% \bottomrule
% % \hline
% \end{tabular}
% \caption{Design operations addressed in this work are distinct from CAD commands in terms of application scenarios, operation diversity and data availability.}
% \label{tab:compare_cad}
% \end{table}

Prior research on parametric modeling has predominantly pursued two complementary paradigms for 3D object representation: CAD command sequences modeling sequential design operations \cite{khan2024cad, ren2022extrudenet, xu2022skexgen, liu2024split, li2022free2cad, seff2021vitruvion}, and analytical primitives defining geometries through mathematical formulations \cite{yu2022capri, yavartanoo20213dias, huang2023learning, chen2020bsp, chen2019bae, liu2022robust, deng2020cvxnet, paschalidou2020learning, paschalidou2019superquadrics, kawana2020neural, paschalidou2021neural}. Within analytical methods, Constructive Solid Geometry (CSG) trees explicitly construct shapes by applying boolean operations (e.g., union, intersection) to primitive solids \cite{kania2020ucsg, yu2022capri, yu2023dualcsg}, whose inherent relational constraints render them particularly suitable for procedural industrial component generation. Complementary efforts have further explored transforming sketches or freehand drawings into parameterized surfaces and curves for mechanical engineering applications \cite{li2022free2cad, seff2021vitruvion}.However, these approaches inherit inherent constraints: Their applicability remains primarily confined to mechanical/industrial components with regular geometries, demonstrating limited generalization capacity for organic or free-form 3D assets, while fundamentally lacking compatibility with standard digital content creation (DCC) workflows due to non-native data representations that impede integration with animation, sculpting, and photorealistic rendering pipelines.

We distinguish the design operations discussed in this chapter from CAD commands, which demonstrate marked divergences across multiple aspects. In terms of application scenarios, design operations are primarily used in industries such as film, gaming, and advertising for modeling scene objects, characters and retail products, mainly utilizing 3D content creation platforms like Blender, Maya, and 3ds Max. In contrast, CAD commands are predominantly employed in industrial manufacturing for mechanical components, primarily through software such as SolidWorks, AutoCAD, and Fusion360. Regarding operational diversity, design operations encompass over 19 commonly used operations integrated in this chapter along with dozens of less frequently used operations. CAD commands mainly rely on Sketching and Extrusion, where Sketching creates closed 2D profiles and Extrusion converts these profiles into 3D solids. Note that the mechanism of CAD Extrusion fundamentally differs from \emph{Extrude} in design operations. Additionally, in terms of data availability, there currently exists no datasets of any scale for design operation sequences. Given the substantial acquisition challenges inherent to this domain, it is expected that such datasets will remain challenging to obtain through conventional collection methods in the foreseeable future. This stands in stark contrast to CAD systems, which currently provide access to millions of publicly available parametric sequence datasets (\emph{e.g.}, ABC \cite{koch2019abc} and DeepCAD \cite{wu2021deepcad}). While numerous studies have successfully leveraged these CAD repositories for reconstructing manufacturing sequences from images or point clouds, research targeting design operation sequences remains entirely unexplored. Consequently, the compounded challenges of operation complexity and acute data scarcity collectively contribute to significantly greater modeling difficulties for design operations compared to CAD commands.

\section{Methods}\label{sec4}

% \subsection{Reconstructing Modeling Operations from 3D Assets}

% Our task is to
% The editability and reusability of models generated through modeling operations further highlight their advantages. These models are typically highly editable, allowing designers to easily modify geometric structures, topology, or parameters. They can also be reused or adapted for different application scenarios. In contrast, models generated by algorithms are often black-box, making direct editing or reuse challenging and requiring complex post-processing, such as mesh reconstruction or topology optimization. This limitation restricts their applicability in industries requiring high precision and normativity, such as engineering and manufacturing, where models generated through modeling operations remain indispensable.

\subsection{Differentiable operation framework}

\begin{table*}[!t]
\centering
\renewcommand{\arraystretch}{1.2} 
\caption{Summary of operations incorporated in our framework, each accompanied with its continuous and discrete parameters. While operations are referenced by nomenclature in Blender, readers may readily identify corresponding functionalities in other 3D content creation software (Maya, 3ds Max, \emph{etc}).}
% \begin{tabular*}{\hsize}{@{}@{\extracolsep{\fill}}lccc@{}}
% \rowcolors{1}{blue!20}
\begin{tabular}{lccc}
\toprule
% \hline

\multicolumn{2}{c}{Operations}    & Continuous Parameters & Discrete Parameters  \\
\toprule
% \hline

& Subdivision  & - & Subdivision Level \\
\hline
&  &Height, & \\
& Extrude & Width, & Target Face Index \\
&  & Orthogonal Angles & \\
\hline
& Bevel & Width & Segment Count \\
\hline
& Boolean & Primitive Affine & Boolean Type, Primitive Type \\
\hline
& Inset & Width & Target Face Index \\
\hline
& VertexDisplace & Displacement Vector & - \\
\hline
& KnifeCut & - & Target Edge Index\\
\hline
& LoopCut & - & Target Edge Ring\\
\hline
& Mirror & - & Mirror Axis \\
\hline
% \midrule
% \hline
& Twist & Factor & Deform Axis \\
& Stretch & Factor & Deform Axis \\
& Bend & Factor & Deform Axis \\
\multirow{-4}{*}{SimpleDeform}& Taper & Factor & Deform Axis \\
\hline
% \midrule
% \hline
& Rotate & Angles & - \\
& Scale & Scale Factor & - \\
\multirow{-3}{*}{GlobalAffine} & Translate & Translate Vector & - \\
\hline
% \midrule
% \hline
& Rotate & Angles & Target Edge Loop \\
& Scale & Scale Factor & Target Edge Loop \\
\multirow{-3}{*}{EdgeLoopAffine} & Translate & Translate Vector & Target Edge Loop \\
\bottomrule
% \hline
\end{tabular}
% \vspace{5pt}

\label{tab:parameters}
\end{table*}
The differentiable operation framework establishes a hierarchical computational graph that reformulates conventional discrete geometric modeling operations into differentiable units, thereby enabling gradient-based optimization of both continuous and discrete parameters. The architecture comprises two core components: differentiable operation nodes and differentiable operation graphs. Each differentiable operation node $ N(\mathbf{x}, \boldsymbol{\theta}, \mathbf{d}) $ operates as a mathematical function that transforms an input geometric state $ \mathbf{x}$ (\emph{e.g.}, mesh vertices $\mathbf{v}$ or topological connections $\mathbf{f}$) into an output state $ \mathbf{y} $. The node parameters are categorized into continuous parameters $ \boldsymbol{\theta} $ (\emph{e.g.}, \emph{Extrude} heights, \emph{Rotate} angles) that reside in real space, and discrete parameters $ \mathbf{d} $ (\emph{e.g.}, \emph{Boolean} primitive types, \emph{Bevel} segment counts) that take discrete values from a finite set. For continuous parameters, differentiability is achieved through chain rule. For instance, a \emph{VertexDisplace} node implements $ \mathbf{v}_\mathbf{y} = \emph{VertexDisplace}(\mathbf{v}_\mathbf{x}, \boldsymbol{\theta}) = \mathbf{v}_\mathbf{x}+\boldsymbol{\theta}$, where $\mathbf{v}_\mathbf{x}, \mathbf{v}_\mathbf{y} \in \mathbb{R}^{L\times3}$ denotes mesh vertices, and $ \boldsymbol{\theta} \in \mathbb{R}^{L\times3} $ controls displacement magnitude, with gradients $ \nabla_{\boldsymbol{\theta}} \mathbf{v}_\mathbf{y} = \mathbf{1}_{L\times 3}$. Discrete parameters employ probabilistic branching mechanisms, where each possible value corresponds to a distinct branch, whose output is weighted by Softmax-transformed coefficients. 
For example, a \emph{Bevel} node supporting $ k $ segment options ($ \mathbf{d} \in \{1,2,...,k\} $) computes parallel branch outputs $ \{\emph{Bevel}(\mathbf{x},\mathbf{d}=1), ..., \emph{Bevel}(\mathbf{x},\mathbf{d}=k)\} $ and weight the output by probabilities $ \mathbf{p} = Softmax(\mathbf{w}) $, where $ \mathbf{w} \in \mathbb{R}^k $ is a learnable parameter.
This formulation preserves differentiability during training while enabling discrete decision decisions through $ \arg\max(\mathbf{p}) $ during inference.
% converts learnable weights $ \mathbf{w} \in \mathbb{R}^k $ to probabilities via $ \mathbf{p} = \text{Softmax}(\mathbf{w}) $, and outputs the expectation $ \mathbf{y} = \sum_{i=1}^k p_i \cdot \text{\emph{Bevel}}_i(\mathbf{x}) $. 

\subsection{Differentiable graph}
The differentiable operation graph $ \mathcal{G} $ is constructed through sequential composition of $ L $ operation nodes. Given an initial geometric state $ \mathbf{x}^0 $ (\emph{e.g.}, a unit cube), the final output state after $ L $ transformations is expressed as:
\begin{equation}
\mathbf{x}^{L+1} = (\Phi^L \circ \Phi^{L-1} \circ \cdots \circ \Phi^0)(\mathbf{x}^0),
\end{equation}
where $ \Phi^l $ denotes the $l$-th node's operation. The output of each node serves as the input of next node. Critically, all operation nodes derive from a base operation set, which is summarized in Tab. \ref{tab:parameters}. The operation set is recursively arranged from $\Phi^0$ through $\Phi^L$. To adaptively optimize the sequence composition, we introduce a gating mechanism which dynamically modulates the contribution of each node. Each node $ \Phi^l $ is associated with a learnable gating weight $ \omega^l \in \mathbb{R} $, transformed into an activation coefficient $ \gamma^l = \sigma(\omega^l) $ via the sigmoid function $\sigma$. 
The output of the $l$-th node becomes:

\begin{equation}
\mathbf{x}^{l+1} = \{\gamma^l \cdot N^l(\mathbf{x}^l, \boldsymbol{\theta}^l, \mathbf{d}^l), (1 - \gamma^l) \cdot \mathbf{x}^l\},
\end{equation}
where $\gamma^l \in [0,1]$ dynamically interpolates between preserving the prior state $\mathbf{x}^l$ and applying the current node's transformation $ N^l$. When $ \gamma^l \to 1 $, node $ N^l $ fully influences downstream computations; when $ \gamma^l \to 0 $, the node is functionally disabled, effectively pruning redundant operations. This mechanism resolves permutation diversity challenges by allowing the network to autonomously suppress suboptimal operations during training.

Each output state $\mathbf{x}^{l+1}$ of an operation node corresponds to a polygon mesh. The graph is optimized end-to-end by minimizing the Chamfer Distance $ \mathcal{L}_{\text{CD}} $ between final outcome meshes and target shape:
\begin{align}
\mathcal{L}_{\text{CD}}(S_1, S_2) &= \sum_{x \in S_1} \min_{y \in S_2} \|x - y\|_2^2 + \sum_{y \in S_2} \min_{x \in S_1} \|y - x\|_2^2 \\ &= \sum_{x \in S_1} D_{S_2}(x) + \sum_{y \in S_2} D_{S_1}(y),
\label{eqn:vallina_cham}
\end{align}
where $ S_1 $ and $ S_2 $ represent point clouds sampled from the final outcome meshes and target shape, respectively. Backpropagation simultaneously updates continuous parameters $ \theta $, discrete branch weights $ \mathbf{w} $, and gating weights $ \omega $. The objective function implicitly encodes design process constraints through two key mechanisms: 1) Multi-stage sequence length optimization, which favors compact operation sequences by staged relaxation of sequence length constraints, and 2) Injection of domain-specific rules, via heavily weighted penalties on invalid parameter combinations (\emph{e.g.}, conflicting geometric operations). This dual-constrained optimization enables our model to approximate ground-truth construction sequences without direct supervision, which is a critical challenge when ground-truth operation sequences are unavailable.

The final operation sequence is extracted by retaining nodes with $ \gamma^l \geq \tau $ and selecting discrete parameters through $ \mathbf{d}^l = \arg\max(\mathbf{p}^l) $, yielding a branch-free operation sequence with deterministic discrete parameters and optimal continuous parameters, \emph{e.g.}, [\emph{Rotate}($\theta$=0.5, axis=x) → \emph{Bevel}(width=0.1, segments=2)]. The differentiable operation framework bridges neural optimization with operational modeling, preserving both geometric precision and editability.

\subsection{Differentiable operations}
Our framework integrates 19 functionally distinct operations that encompass core 3D modeling capabilities, aligning with industry-standard tools in 3D content creation software. We summarize these operations in Tab. \ref{tab:parameters}. The primary challenge encountered in operations differentiation arises from combinatorial explosion caused by discrete parameter values, which generates exponentially branching operation graphs that hinder computation and optimization. This challenge necessitates strategies to reduce or even eliminate the number of branches, especially for those operations which dramatically amplify branching complexity, such as \emph{Subdivision} and \emph{Extrude}. To this end, the following solutions have been proposed: 1) Discrete-to-Continuous Re-parameterization, which involves artificially constructing intermediate states between discrete parameters through interpolation, thereby transforming the discrete selection space into a continuous manifold and enabling smooth gradient flow and transition between discrete values. 2) Differentiable Approximation. For those discrete parameters that are intrinsically resistant to continuous re-parameterization, we implement a differentiable approximation scheme that substitutes the exact operation mechanisms with functionally analogous but branch-free surrogates during early optimization phases. Based on the convergence state of this surrogate, we constrain the discrete parameter search space and seamlessly transition back to the precise operations.

% \begin{figure}[!t]
%     \includegraphics[width=\linewidth]{figures/system.pdf}
%     \caption{Left: Illustration of our method. Right: Functions of Subdivision, Extrude and \emph{Bevel}.
%     }
%     \label{fig:node_diff}
% % \vspace{-0.15in}
% \end{figure}

\begin{figure*}[!t]
    \includegraphics[width=\linewidth]{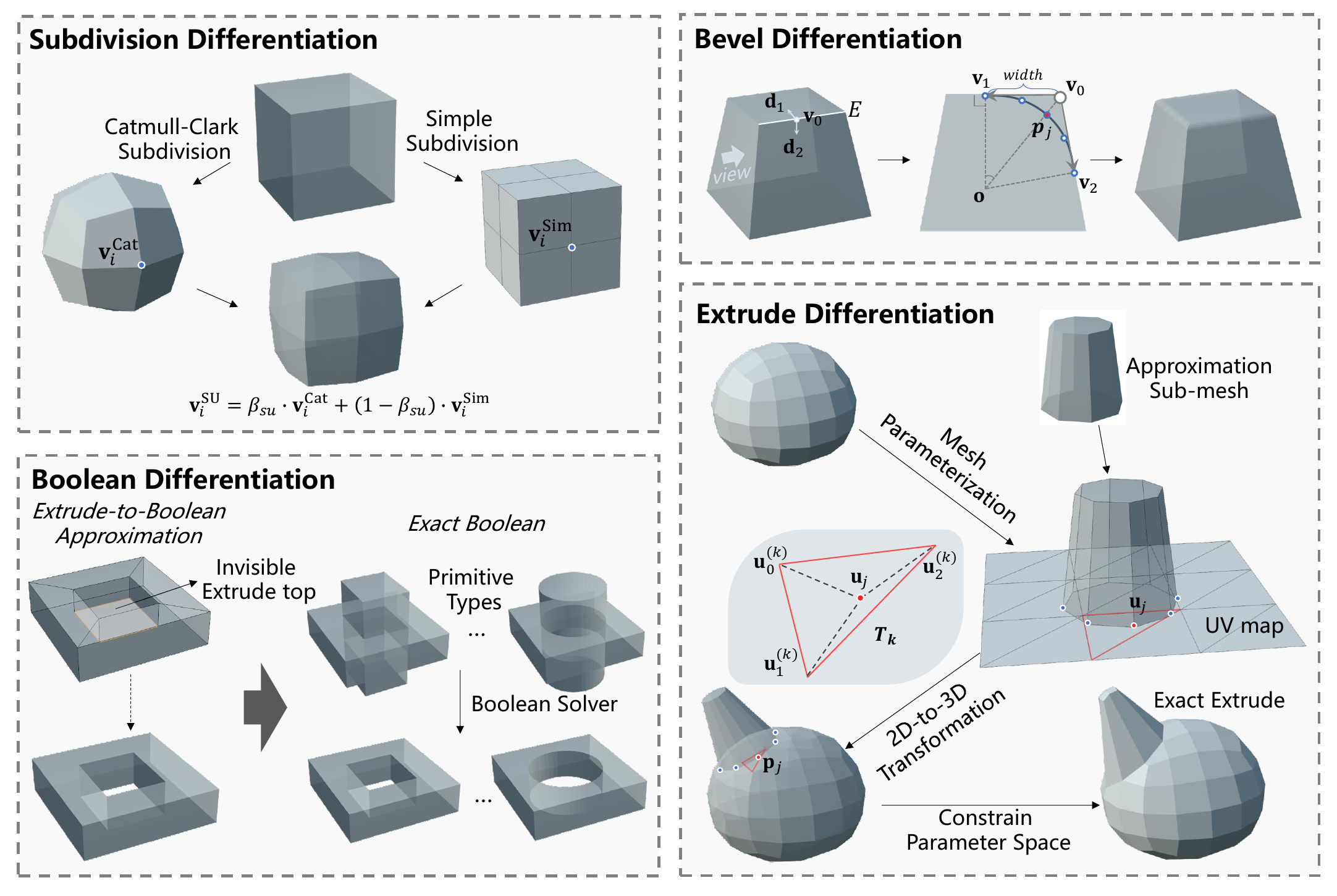}
    \caption{Illustration of differentiation strategies of \emph{Subdivision}, \emph{Bevel}, \emph{Boolean} and \emph{Extrude}. The combinatorial explosion issue caused by discrete parameter is addressed through discrete-to-continuous re-parameterization and geometrical approximation.
    }
    \label{fig:node_diff}
% \vspace{-0.15in}
\end{figure*}

\paragraph{Subdivision.}

The \emph{Subdivision} operation, implementing the Catmull-Clark subdivision algorithm, serves as a fundamental modeling operation for geometric refinement and surface smoothing. It enables hierarchical detail control through recursive mesh partitioning. Combined with other operations like \emph{VertexDisplace} and \emph{Bevel}, as demonstrated in Fig. \ref{fig:node_diff}, it could achieve diverse geometric modifications. The \emph{Subdivision} operation modifies vertex topology by generating new vertices at face centroids and edge midpoints and updates vertex positions according to its neighbors, connected edges and connected faces. While the operation receives a positive integer of subdivision level $I$ as discrete parameter, our framework introduces a dual-path subdivision mechanism that unifies discrete topology manipulation and continuous parameter optimization. It is achieved by parallel process of two types of subdivision schemes: the standard Catmull-Clark subdivision and the so-called simplified subdivision. While standard Catmull-Clark subdivision modifies both geometry and vertex topology, the simplified subdivision alters vertex topology in the same way but preserves original vertex positions. Crucially, both schemes maintain strict element-wise correspondence between their output meshes. Each vertex, edge, and face in the Catmull-Clark subdivision mesh has a topologically equivalent counterpart in the simplified subdivision result. This bijective topological correspondence enables smooth interpolation through a continuous weight parameter $\beta_{su} \in [0,1]$ governing the geometric blend between schemes:
\begin{equation}
\mathbf{v}_i^{\text{SU}} = \beta_{su} \cdot \mathbf{v}_i^{\text{Cat}} + (1-\beta_{su}) \cdot \mathbf{v}_i^{\text{Sim}},
\end{equation}
where $\mathbf{v}_i^{\text{Cat}}$ and $\mathbf{v}_i^{\text{Sim}}$ denote corresponding vertices from Catmull-Clark and simplified subdivisions respectively. The parameter $\beta_{su}$ achieves progressive geometric transitions: $\beta_{su}=1$ yields pure Catmull-Clark refinement with full geometric displacement, while $\beta_{su}=0$ produces pure topological subdivision without surface modification. 

For scenarios where \emph{Subdivision} level is more than one ($I > 1$), we first apply $\lfloor I \rfloor$ full steps of standard Catmull-Clark subdivision. Then, we perform an additional fractional step where we compute the two types of subdivision schemes and perform weighted blending using a weight $\beta_{su}^I = I-\lfloor I \rfloor (\beta_{su}^I \in [0,1))$. This constructs a continuous parameter space, ensuring smooth transitions between integer subdivision levels. The differentiability with respect to both vertex positions and weight parameters is maintained through: 
\begin{equation}
\frac{\partial \mathbf{v}_i^{\text{SU}}}{\partial \beta_{su}} = \mathbf{v}_i^{\text{Cat}} - \mathbf{v}_i^{\text{Sim}}.
\end{equation}
% \begin{equation}
% \frac{\partial \mathbf{v}_i^{final}}{\partial \mathbf{v}_j^{CC}} = w \cdot \delta_{ij}
% \end{equation}
% where $\delta_{ij}$ is the Kronecker delta. 

The continuous parameterization resolves the inherent discontinuity of integer subdivision levels, eliminating the need of branching for each subdivision level. The system could automatically determine optimal subdivision level during optimization,  balancing geometric fidelity against computational complexity.

\paragraph{Extrude}

%This technique finds broad applications in industrial design for constructing structural protrusions (\emph{e.g.}, mechanical flanges), architectural detailing, and organic shape modeling. 
% Conventional \emph{Extrude} workflows suffer from intrinsic differentiability challenges caused by the combinatorial explosion of face selection possibilities, where any combination of single or contiguous faces could theoretically serve as \emph{Extrude} bases.

The \emph{Extrude} operation is a fundamental modeling tool in 3D content creating, widely used across 3D modeling domains. It creates a geometric structure by extending a selected face group along the average normal of the group. \emph{Extrude} operation receives one discrete parameter of target face index and three continuous parameters: \emph{Extrude} height that represents displacement distance along the target face normal, \emph{Extrude} width that represents scaling factor of the target face, and local rotation angle ($\theta$) that specifies in-plane orientation adjustment relative to two orthogonal axes perpendicular to target face normal. The \emph{Extrude} target face might include one face or multiple connected faces of input mesh. Without any constraints, the explosive combination possibility results in a parameter space scaling exponentially with mesh complexity ($O(2^M)$ for an $M$-face input mesh). 
To overcome this limitation, we propose a fully differentiable approximation scheme that decouples geometric transformation from discrete face indexing.

Our solution divides the optimization of \emph{Extrude} in two stages. At the beginning of the first stage, we first apply mesh parameterization on the input mesh, as illustrated in Fig. ~\ref{fig:node_diff}, and obtain a parameterized plane or the so-called UV map $\mathcal{P}$ that is constituted by a set of 2D triangles $\{\mathcal{T}_k\}$ with vertices $\{\mathbf{u}_j^{(k)}\}$. Simultaneously, we apply an approximation sub-mesh, in the shape of a cylinder that is composed of eight interconnected base and top vertices that could move across the surface of the input mesh and imitate the geometric characteristics of \emph{Extrude} operation. The base vertices of the sub-mesh are parameterized via optimizable UV coordinates $\{\mathbf{u}_j\}$ that could moved across the parameterized plane $\mathcal{P}$. At each iteration in first-stage optimization, we establish continuous correspondence between 2D parameter space and 3D geometry. As the vertex coordinates of the input mesh keep changing, we identify the triangle $\mathcal{T}_k$ that contain $\mathbf{u}_j$ and compute its 3D position for each $\mathbf{u}_j$. Specifically, each UV coordinate $\mathbf{u}_j$ dynamically identifies its host triangle $\mathcal{T}_k$ through rasterization: 
\begin{equation}
\mathbf{u}_j \in \mathcal{T}_k \iff \mathcal{E}_i^{(k)}(\mathbf{u}_j) \geq 0 \;\forall i \in \{0,1,2\}, 
\end{equation}
where $\mathcal{E}_i^{(k)}(\mathbf{u}_j) = (\mathbf{u}_j - \mathbf{u}_i^{(k)}) \times (\mathbf{u}_{i+1}^{(k)} - \mathbf{u}_i^{(k)})$ represents the signed edge function that computes the signed distance between the uv coordinate $\mathbf{u}_j$ and a triangle edge connecting two endpoints $\mathbf{u}_i^{(k)}$ and $\mathbf{u}_{i+1}^{(k)}$.
The 3D position $\mathbf{p}_j$ for each $\mathbf{u}_j$ is determined via barycentric interpolation:
\begin{equation}
\mathbf{p}_j = \sum_{i=0}^2 w_i \mathbf{v}_i^{(k)}, \quad w_i = \frac{\text{Area}(\triangle[\mathbf{u}_j, \mathbf{u}_{i+1}^{(k)}, \mathbf{u}_{i+2}^{(k)}])}{\text{Area}(\mathcal{T}_k)},
\label{eqn:bary}
\end{equation}
where $\mathbf{v}_i^{(k)}$ denotes the 3D vertices of $\mathcal{T}_k$ from input mesh, $\triangle[\mathbf{u}_j, \mathbf{v}_{i+1}^{(k)}, \mathbf{v}_{i+2}^{(k)}]$ denotes a 2D triangle form by these three vertices and weights $ w_i$ are computed in 2D parameter space. The analytical computation of $w_i$ in Eqn. \ref{eqn:bary} ensures $C^1$ continuity of the derivative of $w_i$ with respect to $\mathbf{u}_j$. This formulation ensures that moving of $\mathbf{u}_j$ across $\mathcal{P}$ smoothly induces moving of $\mathbf{p}_j$ in 3D space, as the area ratios $w_i$ vary continuously.
Crucially, the derivatives of $\mathbf{p}_j$ with respect to $\mathbf{u}_j$ are preserved through chain rule: %$\frac{\partial \mathbf{p}_j}{\partial \mathbf{u}_j}$
\begin{equation}
\frac{\partial \mathbf{p}_j}{\partial \mathbf{u}_j} = \sum_{i=0}^2 \left( \frac{\partial w_i}{\partial \mathbf{u}_j} \mathbf{v}_i^{(k)} + w_i \frac{\partial \mathbf{v}_i^{(k)}}{\partial \mathbf{u}_j} \right),
\end{equation}
where $\frac{\partial \mathbf{v}_i^{(k)}}{\partial \mathbf{u}_j}$ vanishes since 3D vertices are treated as constants during mesh parameterization. The preserved derivatives enable effective gradient flow from 3D reconstruction errors back to 2D parameter updates.

Except from the base vertices, the approximation sub-mesh share the same continuous parameters as the exact \emph{Extrude} operation, including height $h \in \mathbb{R}^+$, width $w \in \mathbb{R}^+$, and two orthogonal rotation angles $\theta \in [0, 2\pi)$. All of these parameters are jointly optimized at the first stage until convergence. Using the stabilized UV positions, we identify candidate \emph{Extrude} target faces $\mathcal{F}_{\text{candidate}}$. Each face that is covered by the sub-mesh base, and each connected combination of these covered faces, is regarded as one candidate. Each candidate corresponds to one branch. Since the sub-mesh base has been optimized to a relatively proper position, the number of candidates is limited (typically 1-3 candidates). Then we enter the second stage, where the exact \emph{Extrude} operation is applied. 
This strategy, starting with continuous re-parameterization followed by discrete branching, reduces branch complexity from exponential to constant. 

% Upon model convergence, the algorithm executes precise \emph{Extrude} through an approximate \emph{Extrude} sub-mesh process that narrows face index ranges. This final stage establishes individual branches for each face index while retaining three core \emph{Extrude} parameters: height ($h$) controlling normal-aligned displacement, width ($w$) governing lateral scaling, and local rotation angle ($\theta$) specifying in-plane orientation adjustment relative to two orthogonal axes perpendicular to the \emph{Extrude} direction. The preservation of these parameters ensures geometric consistency between the differentiable approximation and final design operations.

\paragraph{Bevel}

The \emph{Bevel} operation is widely employed in mechanical design and geometric modeling to create smooth transitional surfaces between intersecting features, particularly for enhancing structural integrity and visual continuity. Our differentiable implementation provides parametric control over two critical parameters: bevel width $w$ controlling contouring intensity and segment count $K$ controlling smoothing resolution. As illustrated in Fig. \ref{fig:node_diff}, given a target vertex $\mathbf{v}_0$ lying on edge $E$ with adjacent planes $F_1$ and $F_2$, we define vectors orthogonal to edge $E$ and lying within the plane of $F_1$ and $F_2$ as $\mathbf{d}_1$ and $\mathbf{d}_2$ respectively. The geometric construction of \emph{Bevel} operation begins by displacing $\mathbf{v}_0$ along both vectors to obtain translated vertices $\mathbf{v}_1$, $\mathbf{v}_2$:
\begin{equation}
\mathbf{v}_1 = \mathbf{v}_0 + w\frac{\mathbf{d}_1}{\|\mathbf{d}_1\|}, \quad \mathbf{v}_2 = \mathbf{v}_0 + w\frac{\mathbf{d}_2}{\|\mathbf{d}_2\|}.    
\end{equation}

Subsequently, we construct the \emph{Bevel} arc. We first draw a line $L_1$ passing through $\mathbf{v}_1$ and perpendicular to $\mathbf{d}_1$, and determine arc center position $\mathbf{o}$ lying on $L_1$ and arc radius $r=\|\mathbf{v}_1-\mathbf{o}\|$  according to the angle of $\mathbf{d}_1$, $\mathbf{d}_2$. Then we construct orthogonal basis vectors $\mathbf{u}, \mathbf{v}$ in the plane containing $\mathbf{o}$, $\mathbf{v}_1$ and $\mathbf{v}_2$:
\begin{equation}
    \mathbf{u} = \frac{\mathbf{\mathbf{v}_1-\mathbf{o}}}{r}, \quad
    \mathbf{n} = \frac{(\mathbf{v}_1-\mathbf{o}) \times (\mathbf{v}_2-\mathbf{o})}{\|(\mathbf{v}_1-\mathbf{o}) \times (\mathbf{v}_2-\mathbf{o})\|}, \quad
    \mathbf{v} = \mathbf{n} \times \mathbf{u}.
\end{equation}

Arc points $\mathbf{p}(\sigma)$ are then parameterized by angular coordinate $\sigma$:
\begin{equation}
\begin{split}
    &\mathbf{p}(\sigma) = \mathbf{o} + r \left( \cos\sigma \cdot \mathbf{u} + \sin\sigma \cdot \mathbf{v} \right), \\
    & 0 \leq \sigma \leq \sigma_{\text{max}} =\arccos\left( \frac{(\mathbf{v}_1-\mathbf{o}) \cdot (\mathbf{v}_2-\mathbf{o})}{r^2} \right),
\label{eqn:bevel}
\end{split}
\end{equation}
where $\mathbf{p}(0) = \mathbf{v}_1$ and $\mathbf{p}(\sigma_{\text{max}}) = \mathbf{v}_2$. For segment count of $K$, we uniformly sample $K-1$ points $\{\mathbf{p}_j | \mathbf{p}_j = \mathbf{p}(\frac{j\sigma_{\text{max}}}{K}), j=1,...,K-1\}$ along the arc. This construction ensures $C^1$ continuity with respect to arc center coordinates and width parameter due to the analytical formulation in Eqn. \ref{eqn:bevel}. 

Since segment count has weaker geometric impact than bevel width on geometry variation, we fix segments counts $K=5$ during primary optimization to minimize branching complexity. Then we explore discrete options over segments counts $K\in\{1,...,9\}$ to determine the optimal segment count.
% the arc construction involves determining the arc center $O$. We draw two lines $l_1$ and $l_2$ satisfying $l_1: \{P | (P - V_1) \cdot \mathbf{n}_1 = 0\}$ and $l_2: \{P | (P - V_2) \cdot \mathbf{n}_2 = 0\}$, whose intersection point $O$ serves as the arc center satisfying $\|O - V_1\| = \|O - V_2\| = w$. The smooth transition arc is then parameterized as $C(t) = O + w\left[\cos\theta(t)\frac{V_1 - O}{w} + \sin\theta(t)\frac{\mathbf{n}_1 \times \mathbf{n}_2}{\|\mathbf{n}_1 \times \mathbf{n}_2\|}\right]$ where $\theta(t)$ spans the dihedral angle between the planes. 
% To optimize segment counts without premature branching, we initially fix $K=5$ during primary optimization phases, then perform post-convergence discrete optimization over candidate counts $K\in\{1-9\}$ through Eqn. \ref{eqn:cham}. 
% This approach ensures topological stability during early training while enabling precise edge refinement in final stages, maintaining mathematical consistency with our \emph{Boolean} operation framework through shared progressive topology refinement strategies.

\paragraph{Boolean}

The \emph{Boolean} operation is widely used constructive solid geometry operation \cite{zhou2016mesh, kania2020ucsg}, essential for generating high-genus (genus $>$ 1) meshes in 3D modeling. This operation creates new topology through mesh union and mesh difference algorithms between the input mesh and a primitive shape, where mesh union combines the two meshes, and mesh difference cuts out the primitive from input mesh. Direct implementation of \emph{Boolean} operations requires computationally intensive \emph{Boolean} solver, including mesh intersection detection and fragment reconstruction, making it infeasible for iterative training. To address this challenge, we propose a novel approximation scheme based on \emph{Extrude} operation with directional constraints. Specifically, the union process could be approximated by \emph{Extrude} operation of positive \emph{Extrude} height, while the difference process is approximated by \emph{Extrude} operation of negative \emph{Extrude} height, both removing the top of extruded structure, as illustrated in Fig. \ref{fig:node_diff}. 

To create a smooth transition between actual \emph{Extrude} and \emph{Boolean}-approximated \emph{Extrude}, we introduce a \emph{Boolean} weight $\beta_{BL}\in[0,1]$ for each \emph{Extrude} operation to control the transition. This weight regulates the visibility of the extruded structure's top face: higher weight values suppress the top face, while lower ones preserve it. When $\beta_{BL}=1$, the extruded structure degenerates into a planar surface perpendicular to the \emph{Extrude} direction, effectively simulating \emph{Boolean} operation results. The weight is integrated into our loss function. We first divide the Chamfer loss function in Eqn. \ref{eqn:vallina_cham} into two terms, $L_{S_1}$ and $L_{S_2}$, which compute the error of the output results $S_1$ and the target $S_2$ respectively:

\begin{equation}
\mathcal{L}_{\text{CD}}(S_1, S_2) = \sum_{x \in S_1} D_{S_2}(x) + \sum_{y \in S_2} D_{S_1}(y) = L_{S_1} + L_{S_2}.
\label{eqn:cham}
\end{equation}

With regard to the first term $L_{S_1}$, we reweight the errors for those points $x \in T$ sampled from the extruded structure's top face $T$ as:
\begin{equation}
\tilde{L}_{S_1} = \sum_{x \in T} (1-\beta_{BL}) \cdot D_{S_2}(x) + \sum_{x \in S_1\setminus T} D_{S_2}(x).
\label{eqn:reweight_l1}
\end{equation}

When $\beta_{BL}=0$, $\tilde{L}_{S_1}$ is equivalent to $L_{S_1}$, and the operation acts as the actual \emph{Extrude} with top face preserved. When $\beta_{BL}=1$, the points sampled from the extruded structure's top face are excluded from the loss function, where the operation approximates a \emph{Boolean} operation with top face removed. To prevent trivial solutions where \emph{Boolean} weights $\beta_{BL}$ collapse to 1 (inducing minimal loss) in all situations, we further introduce a balancing term in $L_{S_2}$. 
For each point in target $y \in S_2$, we compute its Chamfer Distance in two ways, one against $S_1$ where the extruded structure's top face $T$ is preserved and another against $S_1\setminus T$ where $T$ is removed. We weight these two distances by $\beta_{BL}$ and $1-\beta_{BL}$ as follows:
\begin{equation}
  \tilde{L}_{S_2}  = \sum_{y \in S_2} \left[\beta_{BL} \cdot D_{S_1\setminus T}(y) + (1-\beta_{BL}) \cdot D_{S_1}(y)\right].
\label{eqn:reweight_l2}
\end{equation}

When the target shape $S_2$ requires actual \emph{Extrude} operation instead of \emph{Boolean}, $D_{S_1\setminus T}(y)$ would be larger than $D_{S_1}(y)$ (due to missing geometry), pushing $\beta_{BL}$ toward 0. Conversely, when $S_2$ requires \emph{Boolean}, $D_{S_1}(y)$ would be larger than $D_{S_1\setminus T}(y)$ (due to extra top face), pushing $\beta_{BL} \to 0$. The final loss function combines both $\tilde{L}_{S_1}$ and $\tilde{L}_{S_2}$:
\begin{equation}
\mathcal{L}_{\text{CD}}(S_1, S_2) = \tilde{L}_{S_1} + \tilde{L}_{S_2}.
\label{eqn:boolean_cham}
\end{equation}

In this way, we ensure that for regions where \emph{Boolean}-like geometry is required, $\beta_{BL} \to 1$, while for regions where \emph{Extrude}-like geometry is needed, $\beta_{BL} \to 0$, inducing a smooth geometric transition between the \emph{Extrude} and \emph{Boolean}-approximated results. 
Upon model convergence, \emph{Extrude} operations with $\beta_{BL} \to 1$ are replaced by exact \emph{Boolean} operations parameterized by boolean type (union or difference), primitive type (cube, cylinder and so on) and primitive affine parameter (scaling factor and translation). Since the spatial position of \emph{Boolean} is fixed, \emph{Boolean} solver is executed only once, preserving computational efficiency.

\subsection{User study design for 3D modeling professionals}
To evaluate our algorithm’s impact on human-AI collaboration in 3D design, we conducted a comprehensive user study with 407 qualified 3D modeling professionals. The study adhered to ethical guidelines by implementing anonymized data collection and informed consent agreements explicitly restricting data usage to academic research. We place the design details in the appendix.

\section{Experiments}\label{sec3}

\subsection{Recover design operation sequence}
\begin{figure*}[!ht]
    \includegraphics[width=\linewidth]{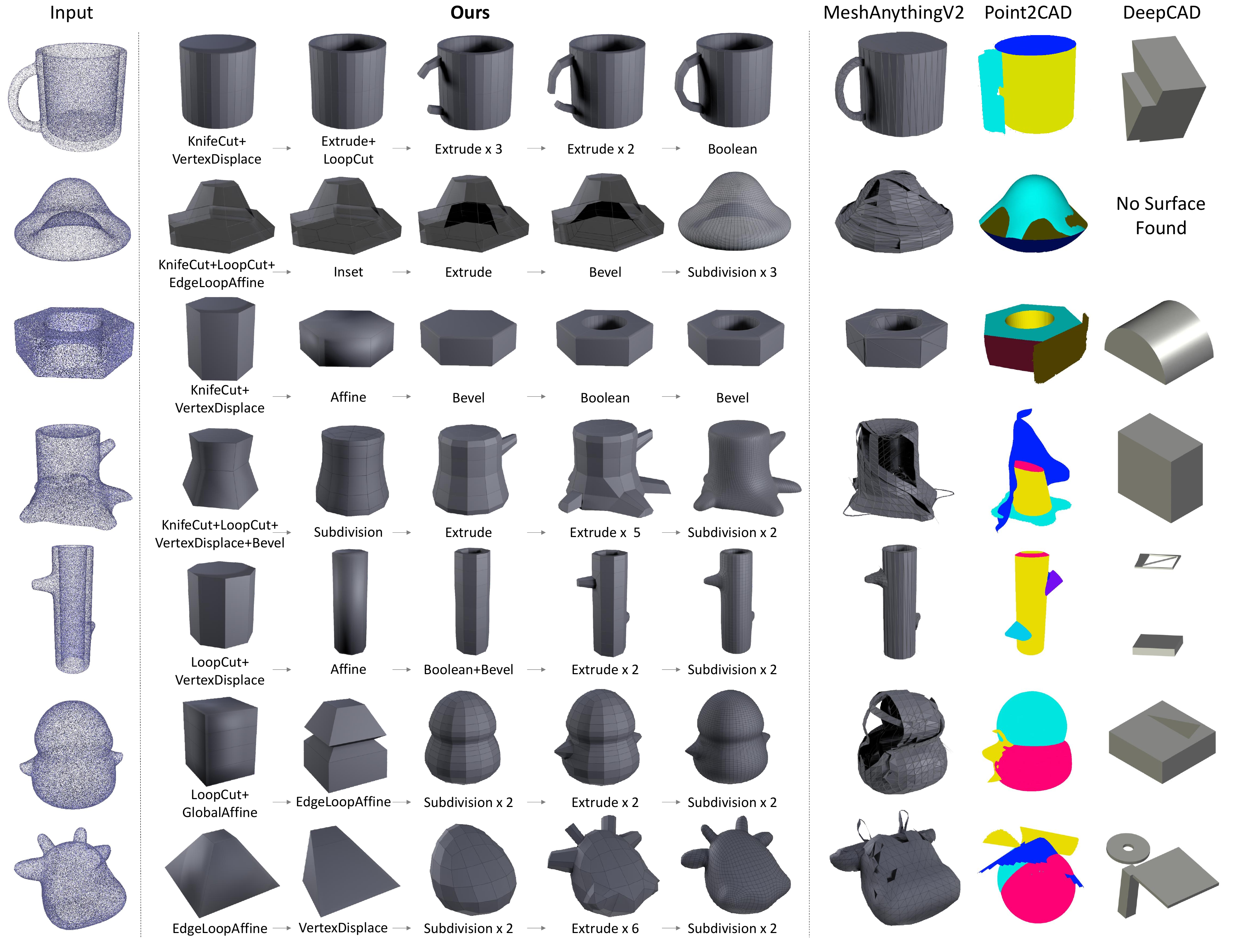}
    \caption{Visualization of our approach and baseline methods. The procedural operation sequence generated by our approach is illustrated in a simplified version, with each operation annotated. This visualization demonstrates that our approach preserves fine details while maintaining topological integrity, unlike other methods which exhibit surface fragmentation or irregular mesh artifacts.}
    \label{fig:compare}
% \vspace{-0.15in}
\end{figure*}

\begin{table*}[!ht]
\renewcommand{\arraystretch}{1.2} 
\centering
% \rowcolors{2}{blue!10}{blue!5}
% \setlength\tabcolsep{1pt}
\caption{Quantitative evaluation on the reconstruction accuracy of our method and baselines (MeshAnythingV2 \cite{chen2024meshanythingv2}, Point2CAD \cite{liu2024point2cad}, DeepCAD \cite{wu2021deepcad}). Our method demonstrates superior geometric accuracy over others with statistically significant differences.}
% \begin{tabular*}{\hsize}{@{}@{\extracolsep{\fill}}lcccc@{}}
\begin{tabular}{lcccc}
\toprule
% \hline

Metrics& \textbf{Ours}    & MeshAnythingV2 & Point2CAD & DeepCAD  \\
\midrule
% \hline
CD ($\times10^{-2}$) $\downarrow$ & \textbf{0.823}$\pm$0.007 & 2.857$\pm$0.121 & 9.349$\pm$1.321 & 30.23$\pm$2.812 \\
Normal Consistency $\uparrow$& \textbf{0.961}$\pm$0.006 & 0.843$\pm$0.032 & 0.757$\pm$0.059 & 0.672$\pm$0.063 \\
$F_1$-score ($\tau$=0.01)  $\uparrow$& \textbf{0.694}$\pm$0.024 & 0.585$\pm$0.033 & 0.503$\pm$0.042 & 0.391$\pm$0.045 \\
$F_1$-score ($\tau$=0.02)  $\uparrow$& \textbf{0.724}$\pm$0.017 & 0.632$\pm$0.028 & 0.558$\pm$0.034 & 0.425$\pm$0.038 \\
\emph{p}-value \emph{vs.} Ours (CD) & - & 0.000005 & 0.000162 & 0.000032 \\
Cohen's \emph{h} \emph{vs.} Ours (CD) & - & -14.5300 & -6.1501 & -9.2775\\
\bottomrule
% \hline
\end{tabular}
% \vspace{5pt}

\label{tab:compare}
\end{table*}

Our method processes both human-designed and AI-generated 3D assets, accepting universal input representations including point clouds and meshes. To the best of our knowledge, this represents the first approach capable of generating design sequences for objects across unlimited categories. To evaluate the effectiveness of our method, we collected 503 3D models from online platforms offering open-access 3D models, including BlenderKit \cite{blenderkit}, Free3D \cite{free3d}, and Sketchfab \cite{sketchfab}. This collection spans diverse application domains such as animation, film, gaming, and e-commerce, encompassing various object types and topological complexities. Given the absence of existing studies directly addressing the identical problem scope, we implemented three functionally relevant baselines for comparison: 1) MeshAnythingV2 \cite{chen2024meshanythingv2} serves as a state-of-the-art reference for generating high-quality meshes with precise geometric alignment, a critical capability for evaluating our shape approximation accuracy and mesh wiring quality; 2) Point2CAD \cite{liu2024point2cad} is a knowledge-driven CAD sequence reconstruction method, providing a benchmark for procedural modeling interpretability, particularly in mechanical domains; 3) DeepCAD \cite{wu2021deepcad} represents data-driven CAD reconstruction methods. While Point2CAD employs manually engineered rules to recover CAD commands, DeepCAD leverages deep learning on large-scale CAD sequences to demonstrate the potential of data-intensive reconstruction. All comparative methods were evaluated across all 503 models using their publicly available implementations with identical point cloud inputs. Each method was run 5 times per model using different random seeds to account for stochasticity.

Quantitative metrics in Tab. \ref{tab:compare} evaluate the reconstruction accuracy using three established measures: 1) Chamfer Distance (CD), which computes bidirectional point-set distances between predictions and ground-truth (lower is better); 2) Normal Consistency, which measures cosine similarity of surface normals to evaluate geometric fidelity (higher is better); 3) $F_1$-score measurements, which combines precision and recall by calculating the percentage of predicted/ground truth points with nearest neighbors within threshold $\tau$ (higher is better). Our approach achieves a mean Chamfer Distance of $0.823\times 10^{-2}$, significantly outperforming MeshAnythingV2 ($2.857\times 10^{-2}$), Point2CAD ($9.349\times 10^{-2}$), and DeepCAD ($30.23\times 10^{-2}$). These differences are statistically significant (p<0.001 for all comparisons), confirming substantial improvements in geometric accuracy. The large magnitude of Cohen’s \emph{h}-values (e.g., -14.53 of MeshAnythingV2 vs. Ours) further highlight the substantial effect size of our method’s improvement. The normal consistency metric also confirms our method's advantage, achieving a score of 0.961 compared to existing methods (0.843-0.672). For the $F_1$-score, our method attains 0.694 ($\tau=0.01$) and 0.724 ($\tau=0.02$), representing improvements of $18.6\%$ and $14.6\%$ over the second-best method at respective thresholds.

Fig. \ref{fig:compare} presents visual comparisons between our method and the baselines. For fair comparison, we rendered mesh outputs from our method and MeshAnythingV2 under identical rendering conditions, while maintaining the default visualization schemes for CAD outputs from Point2CAD and DeepCAD. The visual comparisons demonstrates the distinct advantages of our method in terms of reconstruction precision across diverse shapes. Our approach preserves fine details (\emph{e.g.}, sharp edges and curved contours) while maintaining topological integrity, as evidenced by the absence of surface fragmentation or irregular mesh artifacts observed in other methods. In contrast, alternative methods exhibit critical limitations. While MeshAnythingV2 occasionally generates high-quality meshes (\emph{e.g.}, nuts), it frequently produces incomplete surfaces with holes,  due to its redundant mesh representation. Point2CAD's performance heavily relies on surface pre-segmentation \cite{sharma2020parsenet} and is limited by B-rep flexibility, producing structurally inconsistent outputs. DeepCAD, despite being trained on a large-scale dataset of over 178k CAD samples, shows poor generalization to non-mechanical components and fails to reconstruct valid surfaces in multiple cases.

Fig. \ref{fig:compare} also illustrates the procedural operation sequence derived from our framework, with randomly selected steps simplified for visualization clarity. The decomposition reveals a procedurally coherent modeling process adhering to logical constructive progression, demonstrating how sequential operations systematically generate complex geometry features. 
The illustrative case in the first row demonstrates hierarchical construction of a mug. Starting from a unit cube, it first performs \emph{KnifeCut} and \emph{VertexDisplace} to establish the cylindrical profile of the mug. Subsequent \emph{Extrude} operation vertically creates internal void space, while the \emph{LoopCut} operation bisects side patches along mid-plane, enabling the following \emph{Extrude} to Construct the mug handle by extending from both sides towards the middle. After executing successive \emph{Extrude} operations, a \emph{Boolean} operation is applied at the last step to join the two ends, creating a closed-loop structure. This decomposition aligns with conventional 3D modeling workflows and enables users to trace feature dependencies between intermediate steps and final features.
Crucially, the non-destructive workflow preserves editability at every step. Designers can intervene at any point in the operation sequence to inject original ideation. As evidenced in the mug example, designers retain capacity to modify \emph{Extrude} height, adjust \emph{VertexDisplace} parameters, or insert custom \emph{Bevel} operations without compromising downstream geometry.

\begin{figure*}[!t]
    \includegraphics[width=\linewidth]{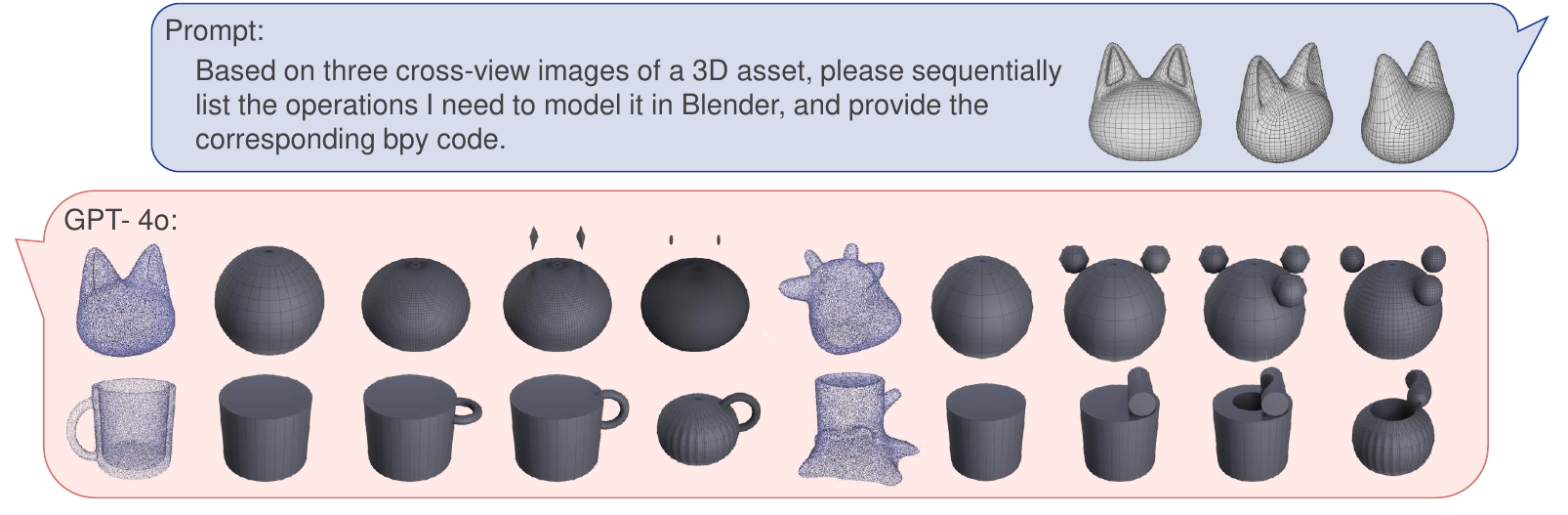}
    \caption{As a baseline comparison, we employed GPT-4o, the state-of-the-art Multimodal Large Language Model (MLLM), to generate design sequences. GPT-4o exhibits repetitive patterns with minimal topological adaptability across distinct geometries and poor geometric consistency between generated outputs and targets.}
    \label{fig:gpt_sequence}
% \vspace{-0.15in}
\end{figure*}

\begin{table*}[!t]
\renewcommand{\arraystretch}{1.2} 
\centering
\caption{We quantitatively validate the alignment of sequences generated by our method against those constructed by professional designers. The results show superior consistency of our method with human-designed workflows.}
% \rowcolors{2}{blue!10}{blue!5}
% \setlength\tabcolsep{1pt}
% \begin{tabular*}{\hsize}{@{}@{\extracolsep{\fill}}lcccc@{}}
\begin{tabular}{lcccc}
\toprule
% \hline
% \rowcolor{blue!20}
Measures & LCS $\uparrow$ & Normalized LCS $\uparrow$ & Levenshtein Distance $\downarrow$ & Levenshtein Similarity $\uparrow$\\

% \hline
\midrule
\textbf{Ours} & \textbf{14.3} & \textbf{0.942} & \textbf{1.4} & \textbf{0.927}\\
GPT-4o & 1.2 & 0.125 & 12.6 & 0.095\\
% \hline
\bottomrule
\end{tabular}

\label{tab:sequence_compare}
\end{table*}

\paragraph{Evaluating Step Rationality.} To validate the alignment between our generated design sequences and those created by human designers, we invited professional designers to manually construct 52 reference models while recording their operations through screen recordings. These recordings were subsequently analyzed to extract the ground-truth design sequences. We quantitatively compare our sequences against the designer sequences using four key metrics that are commonly used in sequence comparison: 
1) Longest Common Subsequence (LCS), which measures the length of the longest subsequence shared by both sequences, reflecting structural similarity in the order of operations while ignoring intermittent differences. Higher is better.
2) Normalized LCS, which is computed as the LCS length divided by the length of the ground-truth sequence, providing a scale-invariant similarity measure ranging from 0 (no overlap between two sequences) to 1 (identical sequences).
3) Levenshtein Distance, which calculates the minimum number of single-operation edits (insertions, deletions, or substitutions) required to transform one sequence (ours) into another (ground-truth), quantifying absolute differences in steps. Lower is better.
4) Levenshtein Similarity, which is derived as 1 - (Levenshtein Distance / ground-truth length), offering a normalized similarity score between 0 (completely different sequences) and 1 (identical sequences).
As a baseline comparison, we employed the state-of-the-art Multimodal Large Language Model (MLLM) GPT-4o to generate design sequences. As illustrated in Fig. \ref{fig:gpt_sequence}, we prompt GPT-4o by providing it with three cross-view images of a 3D model and instruct it to sequentially output Blender modeling operations and corresponding bpy script (Bpy is Blender’s official Python API for programmatic control, enabling mesh creation and Blender operation execution via Python code). In Tab. \ref{tab:sequence_compare}, we compare the quantitative results of our method against GPT-4o-generated sequences. Results are averaged over 52 models containing ground-truth sequences. Our approach achieved an LCS of 14.3 and a Levenshtein Distance of 1.4, indicating that our sequences align with ground-truth sequences by an average of 14.3 ordered operations and require only 1.4 edits per model to match the ground-truth sequences. On normalized metrics, our method attained over $90\%$ alignment in both Normalized LCS and Levenshtein Similarity, demonstrating superior consistency with human-designed workflows. In contrast, GPT-4o exhibited significantly lower alignment across all metrics. 
The operation sequences generated by GPT-4o, visualized in Fig. \ref{fig:gpt_sequence}, expose fundamental limitations. Its sequences exhibit mechanically repetitive patterns and possess minimal topological adaptability across geometrically distinct models. The generated operations are largely limited to elementary operations (\emph{e.g.}, primitive creation, affine transformations, \emph{Boolean} operations), failing to adapt to model-specific requirements. This inflexibility directly leads to poor geometric consistency between its generated models and target models. We attribute this deficiency to MLLMs' inherent inability to understand and organize modeling operations, a specialized modality requiring structural reasoning beyond standard vision-language domains. This observation underscores the necessity of our method's domain-specific design sequence generation paradigm.

\begin{figure*}[!t]
    \includegraphics[width=\linewidth]{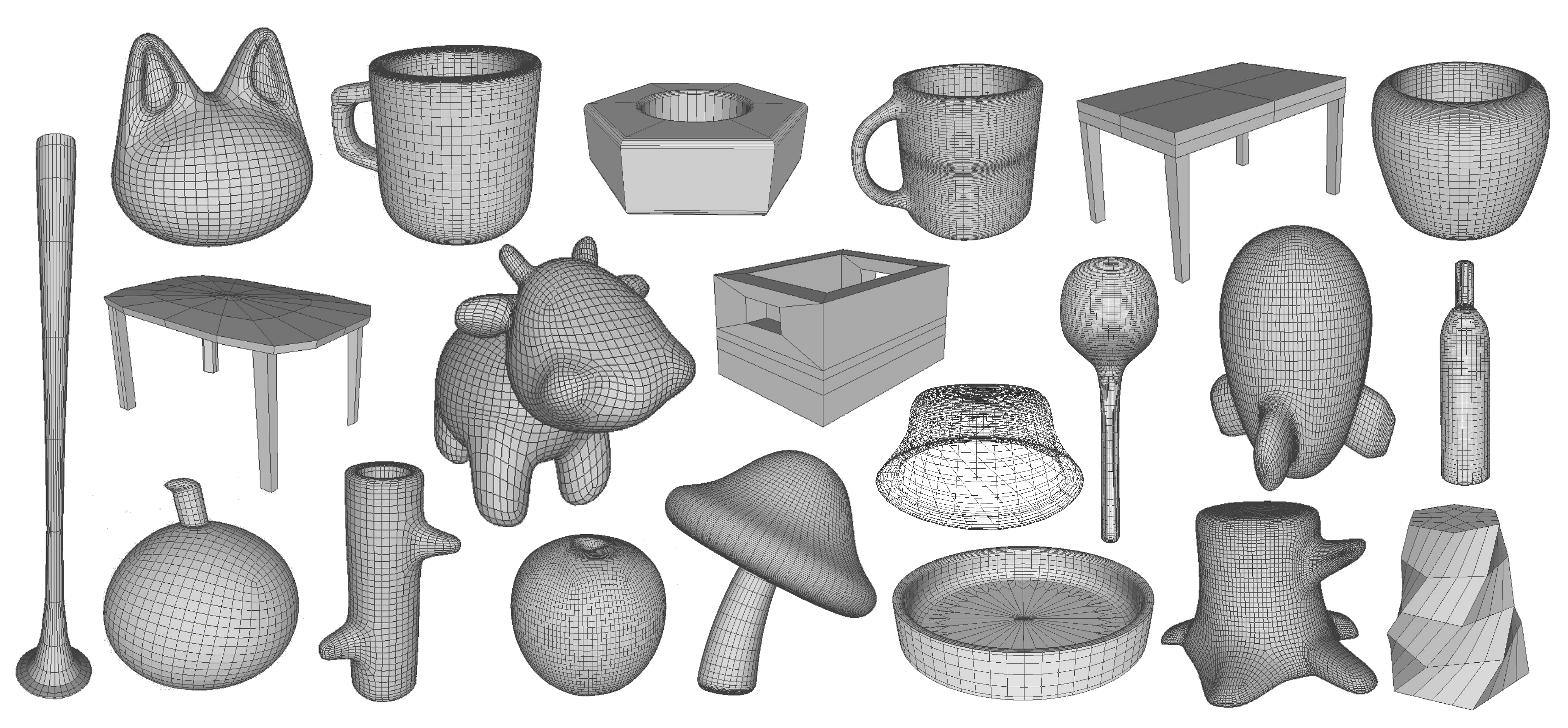}
    \includegraphics[width=\linewidth]{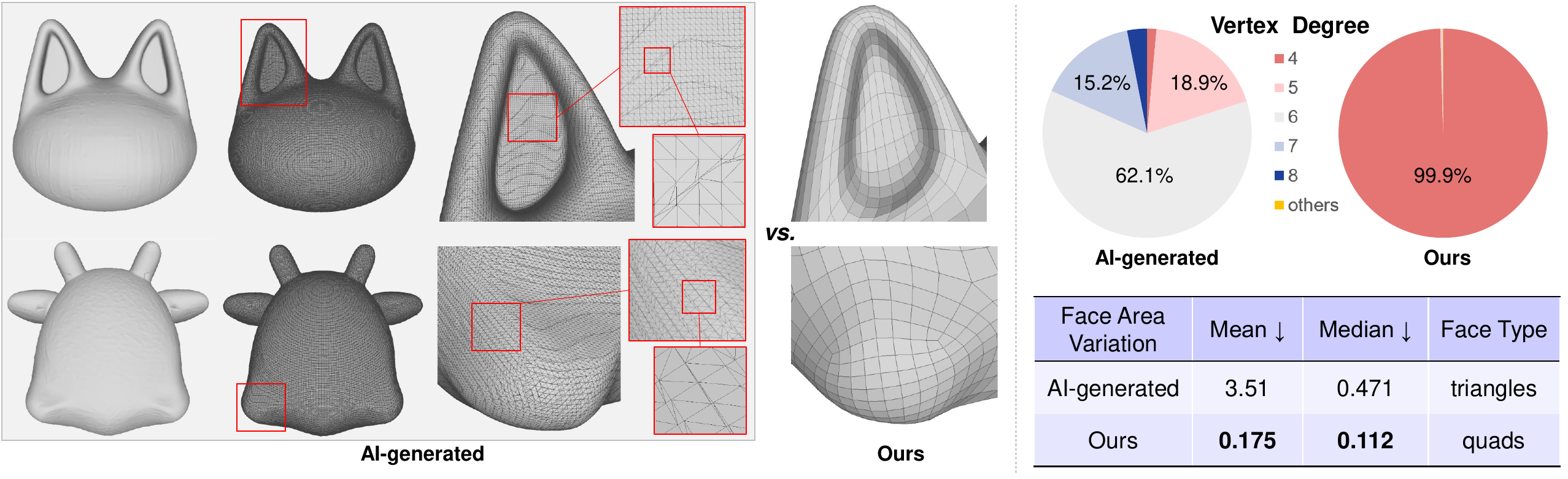}
    \caption{We show the mesh wiring quality of 3D models generated by our framework through design sequence. The generated design sequences create uniformly distributed wiring with little distortion. Selected examples are rendered in transparency mode to visualize their internal structure.}
    \label{fig:wireframe}
% \vspace{-0.15in}
\end{figure*}

\paragraph{Evaluating Mesh Wiring.} Fig. \ref{fig:wireframe} demonstrates the superior mesh wiring quality across diverse 3D models achieved by our method through design sequence. Mesh wiring refers to the structural arrangement of edges and vertices in polygonal meshes, determining both topological integrity and deformation readiness. Our approach generates uniformly distributed meshes with minimized distortion, as evidenced by smooth curvature transitions in organic forms (\emph{e.g.}, the mug's rounded handle) and crisp feature preservation in hard-surface models (\emph{e.g.}, the table legs' sharp edges). The edge flow adheres to the geometric structural features of the model (\emph{e.g.}, aligning with the curvature directions of the ear's concave regions and the protruding contours of the bovine nasal structure in Fig. \ref{fig:wireframe}), thereby enabling precise geometric reconstruction through compact wiring structures and minimizing unnecessary facets crossovers or warping. The optimized wiring topology provides critical advantages including deformation robustness and texture baking efficiency in digital content production. Although our framework does not explicitly incorporate optimization mechanisms tailored for mesh wiring, the operation sequences generated by our framework inherently produce high-quality mesh wiring through compact control that avoids redundancy. These results collectively validate that our operation-based approach fundamentally overcomes the irregular wiring and topological errors inherent in coordinate-based or neural-based representations.
For qualitative comparison, we visualize the meshes generated by current 3D generation methods (implemented with Hunyuan3D-2.0 \cite{zhao2025hunyuan3d}), which demonstrate significant wiring irregularities. These meshes exhibit significant non-uniformity, with some regions containing microscale or highly elongated faces. The edge flow appears disorganized, failing to conform to the model's geometric features (\emph{e.g.}, misaligned with curvature directions of anatomical structures). 
For quantitative comparison, we evaluate the following wiring quality metrics on all 503 models between the two methods: 1) Vertex Degree, which measures the number of edges connected to each vertex, quantifying topological regularity. Lower variance indicates more uniform edge distribution, critical for avoiding mesh artifacts and distortion. 2) Face Area Variation, which computes the standard deviation of adjacent face areas relative to a reference face. Lower values reflect smoother area transitions, essential for maintaining simulation stability and visual coherence. Our meshes exhibit $99.9\%$ vertices with degree 4, whereas AI-generated meshes exhibit irregular vertex degrees (predominantly 5, 6, and 7). High-degree vertices ($\geq$6) act as topological singularities during subdivision or deformation, inducing stress concentrations that lead to mesh tearing. Besides, our meshes demonstrate much lower face area variation compared to AI methods (0.175 vs. 3.51 for mean value and 0.112 vs. 0.471 for median value), which proves superior wiring uniformity. Besides, our method generates quad meshes that inherently maintain topological controllability and geometric smoothness during subdivision, deformation, and UV unwrapping, which are critical properties in designers' workflows. In contrast, the triangular meshes generated by AI methods are not the commonly used form in designers' workflows.

\subsection{Edit generated operations}
\begin{figure*}[!t]
    \includegraphics[width=\linewidth]{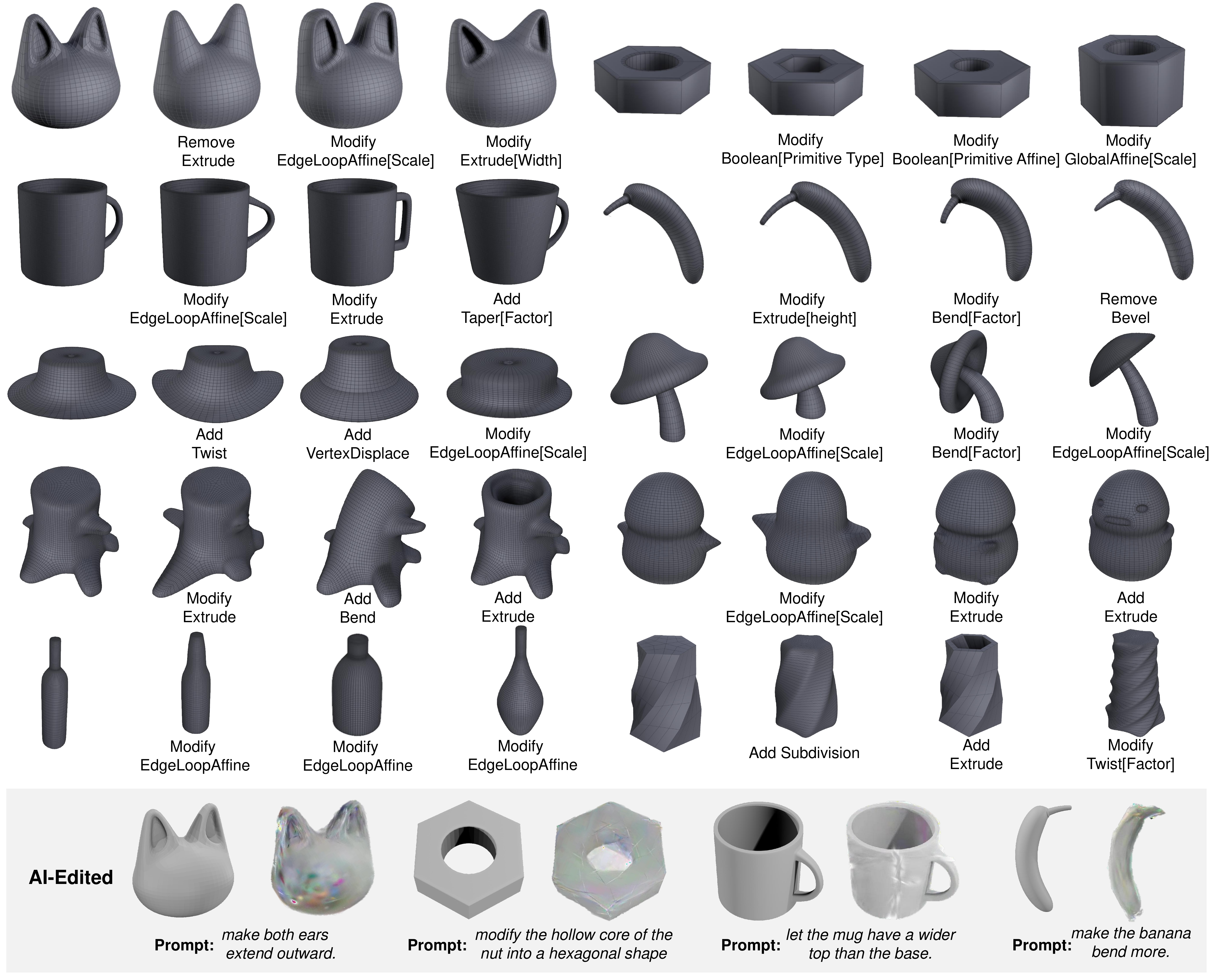}
    \caption{The design sequences generated by our method align with designers' typical workflows and preserves editability at every step, enabling granular manipulation via parameter adjustment, operation insertion, or operation removal. In contrast, GaussianEditor, a SOTA MLLM-based 3D editing method, exhibits critical limitations in precise model editing.
    }
    \label{fig:modify}
    
% \vspace{-0.15in}
\end{figure*}
\begin{figure*}[!t]
    \includegraphics[width=\linewidth]{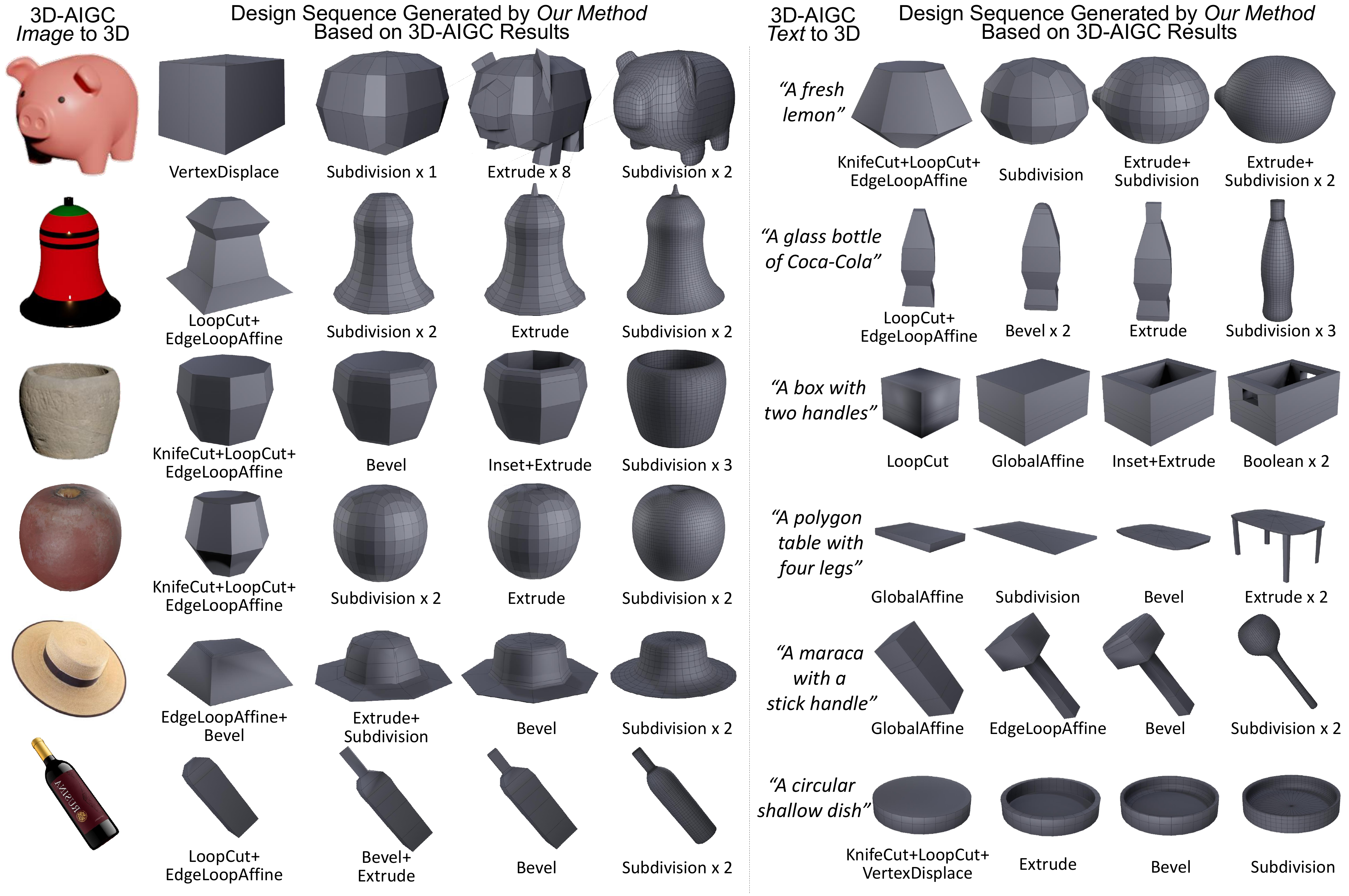}
    \caption{Our framework could be integrated with multimodal 3D generation models, enabling design sequence generation from image/text prompts. 
    }
    \label{fig:llmgen}
% \vspace{-0.15in}
\end{figure*}

Our method generates design sequences that align with designers' typical workflows in 3D content creation software. This intrinsic compatibility enables direct manipulation of operation sequences through parameter adjustment, operation insertion, or operation removal, actions that directly correspond to standard design editing paradigms. Fig. \ref{fig:modify} demonstrates cases where users modify critical parameters while preserving fundamental geometric validity. 
As evidenced by the cat model (Row 1, Left), modifying the \emph{Extrude} parameters enables precise control over feature protrusion, while adjusting \emph{Scale} of \emph{EdgeLoopAffine} reshapes topological contours. The mug model (Row 2, Left) exemplifies combined operation manipulation. Modification of multiple \emph{Extrude} operations converts the curved handle into a rectangular one. \emph{Taper} parameter editing produces smooth volumetric transitions. In the banana model (Row 2, Right), modifying the angle of \emph{Bend} operation introduces curvature deformation, and removal of \emph{Bevel} modifies edge sharpness. 
Each operation preserves granular editability while maintaining mesh integrity, enabling designers to focus on creative exploration rather than geometric validation and suggesting broad applicability in industrial design and animation pipelines where iterative model variations are essential. Besides, this flexibility ensures that even suboptimal outputs serve as fully editable templates for refinement, substantially enhancing human-AI collaboration efficiency. 
As a comparative baseline, we apply the state-of-the-art MLLM-based 3D editing method (GaussianEditor \cite{chen2024gaussianeditor}) for text-driven editing of 3D models (Fig. \ref{fig:modify}, bottom). While current 3D generation methods excel at synthesizing geometrically plausible models from scratch, they exhibit critical limitations in precise model editing: the edited outputs show not only semantic misalignment with input specifications (\emph{e.g.}, failing to modify the nut's hollow core shape of the nut) but also global distortion to the entire 3D model.

\subsection{Interact with Image/Text-to-3D models }

Our framework demonstrates intrinsic compatibility with multimodal 3D generation architectures. The workflow starts with generating 3D meshes from image or text prompts using a 3D generative model (specifically utilizing Hunyuan3D-2.0 \cite{zhao2025hunyuan3d} in our implementation). Point clouds sampled from these meshes then serve as input to our framework. As shown in Fig. \ref{fig:llmgen}, this integration enables our method to successfully recover logically structured design sequences from multimodal inputs while preserving both geometric fidelity and editable parametric controls. 
The left column shows results generated from images prompts. In the cartoon pig model (Row 1, Left), our method recovers fine structures including the nose, two ears and four legs, each with separate \emph{Extrude} operation, suggesting that that are all independently editable. Notably, even the detailed protrusion at the top of jingle bell model (Row 2, Left) is preserved. The right column shows results generated from text prompts. In the lemon model (Row 1, Right), the main body was first constructed using three operations. A \emph{Subdivision} operation established the topological foundation for two subsequent \emph{Extrude} operations that formed the head and tail protrusions, followed by another \emph{Subdivision} to create the final smoothed surface. The coca-cola model (Row 2, Right) exhibits optimal curvilinear profile continuity through gradient-based parameter optimization within our differentiable operational framework. In the box model (Row 3, Right), two \emph{Boolean} operations form two handles, producing a genus-2 topological structure.
This multimodal integration significantly extends the applicability of our framework across diverse 3D content creation scenarios.

\subsection{User study}
% \begin{figure}[!t]
%     \includegraphics[width=\linewidth]{figures/survey_0.pdf}
%     \caption{We conduct a survey among 407 professional 3D designers to assess needs and pains of contemporary 3D-AIGC technology and evaluate the efficacy of our proposed methodology. 
%     }
%     \label{fig:survey_0}
    
% \end{figure}
\begin{figure*}[!t]
    \includegraphics[width=\linewidth]{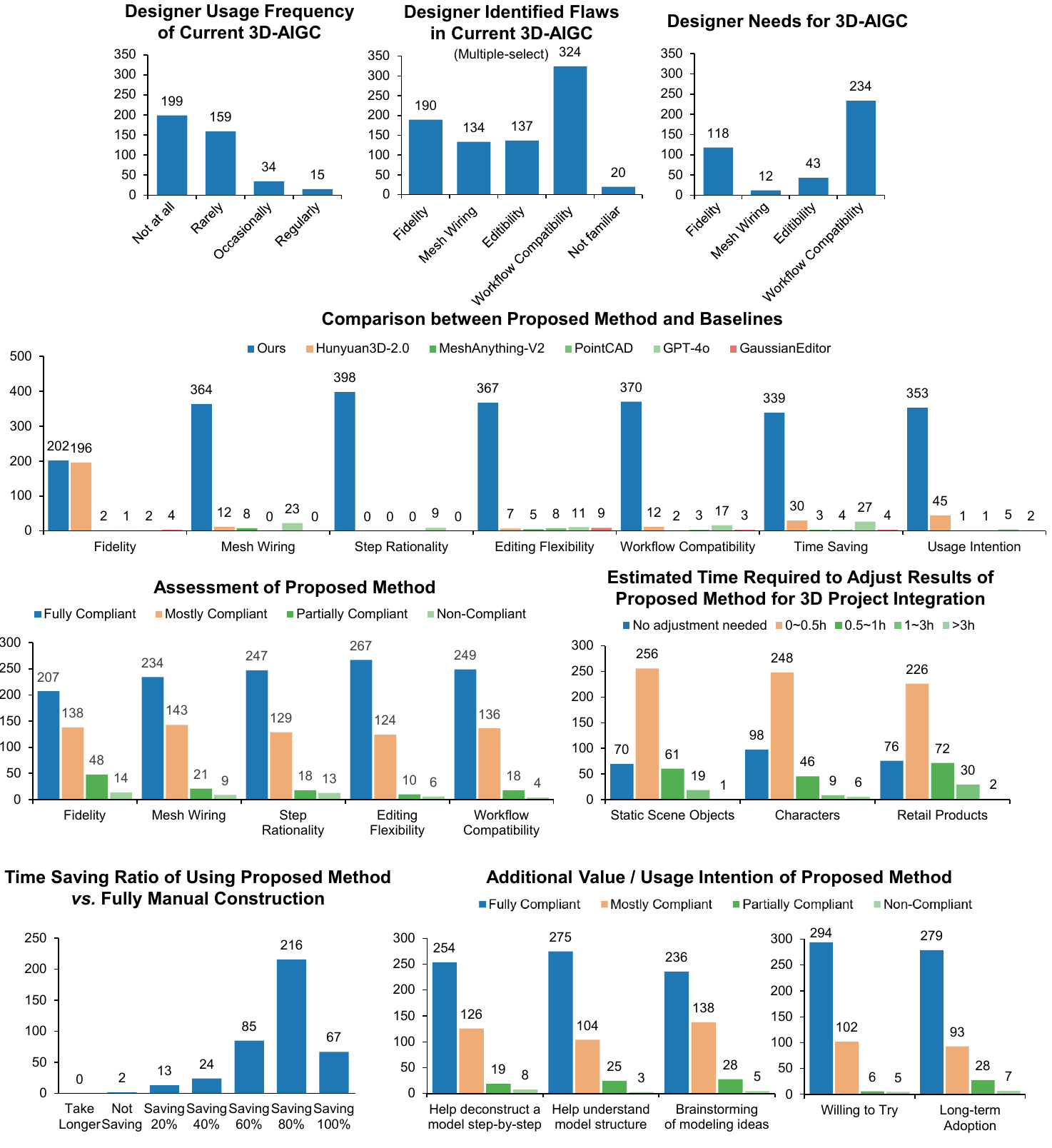}
    \caption{We conducted a comprehensive user study with 407 professional 3D designers to identify challenges and limitations in contemporary 3D-AIGC technology workflows and evaluate the efficacy of our proposed methodology.
    }
    \label{fig:survey}
    
\end{figure*}
%Our survey revealed strong endorsement of our method, with 91.4\% of participants expressing high intent to adopt it in their workflows on a long-term basis.
To evaluate our algorithm’s impact on human-AI collaboration in 3D design, we conducted a comprehensive questionnaire-based user study with 407 qualified 3D modeling professionals recruited via industry forums, networks and academic collaborations. The survey instrument is composed of five interconnected modules with 31 closed and open-ended items. Module 1 captured participant background for segmentation. Module 2 identified pain points in current tools (e.g., fidelity, mesh topology, editability) and quantified time savings estimates across object types (static objects, characters, products). Module 3 performed a comparative analysis against key baselines across seven criteria including fidelity, mesh wiring, step rationality, and workflow compatibility, revealing automation-usability trade-offs. Module 4 specifically rated the proposed method's compliance with core requirements and time efficiency. Module 5 gathered subjective feedback on procedural decomposition impact, adoption willingness, and workflow enhancement potential. Employing cognitive pilot testing and expert validation ensured methodological robustness. Response distributions are visualized in Fig. \ref{fig:survey_intro} and Fig. \ref{fig:survey}.

% To rigorously evaluate our method’s efficacy in enhancing 3D human-AI collaboration, we conducted a user study with 407 professional 3D designers representing multiple sectors including gaming, film production, animation, interior design, and digital art. Response distributions are visualized in Fig. \ref{fig:survey_intro} and Fig. \ref{fig:survey}.

Our study reveals that while 95.1\% of participants reported familiarity with existing 3D generative tools such as Hunyuan3D-2.0, their usage frequency remains low, with only 12\% reporting occasional or regular use. When queried about the reasons for non-adoption, participants predominantly (79.6\%) attributed this disconnect to poor integration with existing workflows, which was also identified as the most critical requirement by design professionals. Specifically, 70\% of practitioners spent over 3 hours modifying 3D-AIGC results for production integration. Moreover, more than 72\% observed no time savings (or even increased time costs) compared to manual modeling from scratch. These findings collectively highlight the fundamental necessity for workflow-compatible solutions that bridge machine automation and human-centric design paradigms.

When comparing our method to existing solutions, participants showed a significantly greater preference for our approach over existing solutions. Specifically, our method outperformed alternatives by substantial margins (achieving >80\% preference) across six dimensions: mesh wiring, step rationality, editing flexibility, workflow compatibility, time saving and usage intention. Only for geometric fidelity did our method attain comparable voting shares to Hunyuan3D-2.0 (49.6\% \emph{vs.} 48.2\%).

Detailed assessment of our method demonstrates strong user approval across all assessed dimensions. Regarding geometric fidelity, 84.8\% of designers completely or mostly agreed that our approach maintains high shape fidelity between input specifications and output geometries. For mesh wiring, 92.6\% praised the structural coherence and logical connectivity. For step rationality, 92.4\% of participants confirm that the modeling decomposition logic aligns with conventional design process. For editing flexibility, 96.1\% acknowledged the editing flexibility, \emph{i.e.}, the granular editability over operation sequences. Additionally, 94.6\% are fully or mostly agreed that the proposed method is compatible with their workflow. These findings align with our core design philosophy of balancing automation with manual adjustability, which is absent in current end-to-end generative systems. Regarding workflow efficiency, participants widely recognized significant efficiency improvements. Participants were asked about adjustment time needed to integrate the results of the proposed method into 3D projects across three core use cases: static scene object modeling, character modeling, and retail product design. On average, over 75\% of participants reported less than 0.5 hours required for project integration of our method. Besides, almost all participants acknowledged that our method saves time in their 3D projects. 90.4\% of participants reported more than 60\% time savings compared to fully manual modeling, with 69.5\% indicating more than 80\% time savings. 

Beyond core functionality, participants highlighted additional value. 93.4\% (fully or mostly compliant) of participants cited its benefit in the step-by-step deconstruction and visualization of complex models, with 93.1\% acknowledging its utility of helping users to understand the model structure. Besides, 91.9\% of participants acknowledged its utility in facilitating brainstorming and accelerating modeling ideation.
Crucially, 97.3\% (fully or mostly compliant) indicated willingness to explore the proposed method, while 91.4\% expressed desire for long-term integration into daily workflows.

\section{Conclusion}
In this study, we translate arbitrary 3D assets into editable design sequences (\emph{e.g.}, Extrude, Boolean) via a novel differentiable operation graph. By reformulating 19 modeling operations into gradient-optimizable units, our method jointly optimizes continuous and discrete parameters without ground-truth sequence supervision. Our approach aligns with designer workflows resulting in substantial time savings, enabling parametric editability and integration with multimodal 3D-AIGC systems, validated by quantitative and qualitative evaluation. This research bridges AI automation with professional designer workflows, paving the way for more collaborative and efficient 3D content creation.

\bibliographystyle{IEEEtrans}% specify bibliography style
\bibliography{ieee-bibliography}% common bib file

% Generated by IEEEtran.bst, version: 1.14 (2015/08/26)
\begin{thebibliography}{10}
\providecommand{\url}[1]{#1}
\csname url@samestyle\endcsname
\providecommand{\newblock}{\relax}
\providecommand{\bibinfo}[2]{#2}
\providecommand{\BIBentrySTDinterwordspacing}{\spaceskip=0pt\relax}
\providecommand{\BIBentryALTinterwordstretchfactor}{4}
\providecommand{\BIBentryALTinterwordspacing}{\spaceskip=\fontdimen2\font plus
\BIBentryALTinterwordstretchfactor\fontdimen3\font minus \fontdimen4\font\relax}
\providecommand{\BIBforeignlanguage}[2]{{%
\expandafter\ifx\csname l@#1\endcsname\relax
\typeout{** WARNING: IEEEtran.bst: No hyphenation pattern has been}%
\typeout{** loaded for the language `#1'. Using the pattern for}%
\typeout{** the default language instead.}%
\else
\language=\csname l@#1\endcsname
\fi
#2}}
\providecommand{\BIBdecl}{\relax}
\BIBdecl

\bibitem{lee2024dreamflow}
K.~Lee, K.~Sohn, and J.~Shin, ``Dreamflow: High-quality text-to-3d generation by approximating probability flow,'' in \emph{12th International Conference on Learning Representations, ICLR 2024}, 2024.

\bibitem{babu2024hyperfields}
S.~Babu, R.~Liu, A.~Zhou, M.~Maire, G.~Shakhnarovich, and R.~Hanocka, ``Hyperfields: Towards zero-shot generation of nerfs from text,'' in \emph{International Conference on Machine Learning}.\hskip 1em plus 0.5em minus 0.4em\relax PMLR, 2024, pp. 2230--2247.

\bibitem{jones2023shapecoder}
R.~K. Jones, P.~Guerrero, N.~J. Mitra, and D.~Ritchie, ``Shapecoder: Discovering abstractions for visual programs from unstructured primitives,'' \emph{ACM Transactions on Graphics (TOG)}, 2023.

\bibitem{jones2021shapemod}
R.~K. Jones, D.~Charatan, P.~Guerrero, N.~J. Mitra, and D.~Ritchie, ``Shapemod: macro operation discovery for 3d shape programs,'' \emph{ACM Transactions on Graphics (TOG)}, vol.~40, no.~4, pp. 1--16, 2021.

\bibitem{jones2020shapeassembly}
R.~K. Jones, T.~Barton, X.~Xu, K.~Wang, E.~Jiang, P.~Guerrero, N.~J. Mitra, and D.~Ritchie, ``Shapeassembly: Learning to generate programs for 3d shape structure synthesis,'' \emph{ACM Transactions on Graphics (TOG)}, vol.~39, no.~6, pp. 1--20, 2020.

\bibitem{raistrick2023infinite}
A.~Raistrick, L.~Lipson, Z.~Ma, L.~Mei, M.~Wang, Y.~Zuo, K.~Kayan, H.~Wen, B.~Han, Y.~Wang \emph{et~al.}, ``Infinite photorealistic worlds using procedural generation,'' in \emph{Proceedings of the IEEE/CVF conference on computer vision and pattern recognition}, 2023, pp. 12\,630--12\,641.

\bibitem{raistrick2024infinigen}
A.~Raistrick, L.~Mei, K.~Kayan, D.~Yan, Y.~Zuo, B.~Han, H.~Wen, M.~Parakh, S.~Alexandropoulos, L.~Lipson \emph{et~al.}, ``Infinigen indoors: Photorealistic indoor scenes using procedural generation,'' in \emph{Proceedings of the IEEE/CVF Conference on Computer Vision and Pattern Recognition}, 2024, pp. 21\,783--21\,794.

\bibitem{huang2024blenderalchemy}
I.~Huang, G.~Yang, and L.~Guibas, ``Blenderalchemy: Editing 3d graphics with vision-language models,'' in \emph{European Conference on Computer Vision}.\hskip 1em plus 0.5em minus 0.4em\relax Springer, 2024, pp. 297--314.

\bibitem{gu2025blendergym}
Y.~Gu, I.~Huang, J.~Je, G.~Yang, and L.~Guibas, ``Blendergym: Benchmarking foundational model systems for graphics editing,'' \emph{arXiv preprint arXiv:2504.01786}, 2025.

\bibitem{chen2024meshanything}
Y.~Chen, T.~He, D.~Huang, W.~Ye, S.~Chen, J.~Tang, X.~Chen, Z.~Cai, L.~Yang, G.~Yu \emph{et~al.}, ``Meshanything: Artist-created mesh generation with autoregressive transformers,'' \emph{arXiv preprint arXiv:2406.10163}, 2024.

\bibitem{chen2024meshanythingv2}
Y.~Chen, Y.~Wang, Y.~Luo, Z.~Wang, Z.~Chen, J.~Zhu, C.~Zhang, and G.~Lin, ``Meshanything v2: Artist-created mesh generation with adjacent mesh tokenization,'' \emph{arXiv preprint arXiv:2408.02555}, 2024.

\bibitem{siddiqui2024meshgpt}
Y.~Siddiqui, A.~Alliegro, A.~Artemov, T.~Tommasi, D.~Sirigatti, V.~Rosov, A.~Dai, and M.~Nie{\ss}ner, ``Meshgpt: Generating triangle meshes with decoder-only transformers,'' in \emph{Proceedings of the IEEE/CVF conference on computer vision and pattern recognition}, 2024, pp. 19\,615--19\,625.

\bibitem{mildenhall2021nerf}
B.~Mildenhall, P.~P. Srinivasan, M.~Tancik, J.~T. Barron, R.~Ramamoorthi, and R.~Ng, ``Nerf: Representing scenes as neural radiance fields for view synthesis,'' \emph{Communications of the ACM}, vol.~65, no.~1, pp. 99--106, 2021.

\bibitem{pooledreamfusion}
B.~Poole, A.~Jain, J.~T. Barron, and B.~Mildenhall, ``Dreamfusion: Text-to-3d using 2d diffusion,'' in \emph{The Eleventh International Conference on Learning Representations}.

\bibitem{kerbl20233d}
B.~Kerbl, G.~Kopanas, T.~Leimk{\"u}hler, and G.~Drettakis, ``3d gaussian splatting for real-time radiance field rendering.'' \emph{ACM Trans. Graph.}, vol.~42, no.~4, pp. 139--1, 2023.

\bibitem{yi2023gaussiandreamer}
T.~Yi, J.~Fang, J.~Wang, G.~Wu, L.~Xie, X.~Zhang, W.~Liu, Q.~Tian, and X.~Wang, ``Gaussiandreamer: Fast generation from text to 3d gaussians by bridging 2d and 3d diffusion models,'' in \emph{CVPR}, 2024.

\bibitem{khan2024cad}
M.~S. Khan, E.~Dupont, S.~A. Ali, K.~Cherenkova, A.~Kacem, and D.~Aouada, ``Cad-signet: Cad language inference from point clouds using layer-wise sketch instance guided attention,'' in \emph{Proceedings of the IEEE/CVF Conference on Computer Vision and Pattern Recognition}, 2024, pp. 4713--4722.

\bibitem{ren2022extrudenet}
D.~Ren, J.~Zheng, J.~Cai, J.~Li, and J.~Zhang, ``Extrudenet: Unsupervised inverse sketch-and-extrude for shape parsing,'' in \emph{European Conference on Computer Vision}.\hskip 1em plus 0.5em minus 0.4em\relax Springer, 2022, pp. 482--498.

\bibitem{xu2022skexgen}
X.~Xu, K.~D. Willis, J.~G. Lambourne, C.-Y. Cheng, P.~K. Jayaraman, and Y.~Furukawa, ``Skexgen: Autoregressive generation of cad construction sequences with disentangled codebooks,'' in \emph{International Conference on Machine Learning}.\hskip 1em plus 0.5em minus 0.4em\relax PMLR, 2022, pp. 24\,698--24\,724.

\bibitem{liu2024split}
Y.~Liu, J.~Chen, S.~Pan, D.~Cohen-Or, H.~Zhang, and H.~Huang, ``Split-and-fit: Learning b-reps via structure-aware voronoi partitioning,'' \emph{ACM Transactions on Graphics (TOG)}, vol.~43, no.~4, pp. 1--13, 2024.

\bibitem{li2022free2cad}
C.~Li, H.~Pan, A.~Bousseau, and N.~J. Mitra, ``Free2cad: Parsing freehand drawings into cad commands,'' \emph{ACM Transactions on Graphics (TOG)}, vol.~41, no.~4, pp. 1--16, 2022.

\bibitem{seff2021vitruvion}
A.~Seff, W.~Zhou, N.~Richardson, and R.~P. Adams, ``Vitruvion: A generative model of parametric cad sketches,'' in \emph{International Conference on Learning Representations}, 2021.

\bibitem{yu2022capri}
F.~Yu, Z.~Chen, M.~Li, A.~Sanghi, H.~Shayani, A.~Mahdavi-Amiri, and H.~Zhang, ``Capri-net: Learning compact cad shapes with adaptive primitive assembly,'' in \emph{Proceedings of the IEEE/CVF Conference on Computer Vision and Pattern Recognition}, 2022, pp. 11\,768--11\,778.

\bibitem{yavartanoo20213dias}
M.~Yavartanoo, J.~Chung, R.~Neshatavar, and K.~M. Lee, ``3dias: 3d shape reconstruction with implicit algebraic surfaces,'' in \emph{Proceedings of the IEEE/CVF International Conference on Computer Vision}, 2021, pp. 12\,446--12\,455.

\bibitem{huang2023learning}
X.~Huang, Y.~Zhang, K.~Chen, T.~Li, W.~Zhang, and B.~Ni, ``Learning shape primitives via implicit convexity regularization,'' in \emph{Proceedings of the IEEE/CVF International Conference on Computer Vision}, 2023, pp. 3642--3651.

\bibitem{chen2020bsp}
Z.~Chen, A.~Tagliasacchi, and H.~Zhang, ``Bsp-net: Generating compact meshes via binary space partitioning,'' in \emph{Proceedings of the IEEE/CVF Conference on Computer Vision and Pattern Recognition}, 2020, pp. 45--54.

\bibitem{chen2019bae}
Z.~Chen, K.~Yin, M.~Fisher, S.~Chaudhuri, and H.~Zhang, ``Bae-net: Branched autoencoder for shape co-segmentation,'' in \emph{Proceedings of the IEEE/CVF International Conference on Computer Vision}, 2019, pp. 8490--8499.

\bibitem{liu2022robust}
W.~Liu, Y.~Wu, S.~Ruan, and G.~S. Chirikjian, ``Robust and accurate superquadric recovery: a probabilistic approach,'' in \emph{Proceedings of the IEEE/CVF Conference on Computer Vision and Pattern Recognition}, 2022, pp. 2676--2685.

\bibitem{deng2020cvxnet}
B.~Deng, K.~Genova, S.~Yazdani, S.~Bouaziz, G.~Hinton, and A.~Tagliasacchi, ``Cvxnet: Learnable convex decomposition,'' in \emph{Proceedings of the IEEE/CVF Conference on Computer Vision and Pattern Recognition}, 2020, pp. 31--44.

\bibitem{paschalidou2020learning}
D.~Paschalidou, L.~V. Gool, and A.~Geiger, ``Learning unsupervised hierarchical part decomposition of 3d objects from a single rgb image,'' in \emph{Proceedings of the IEEE/CVF Conference on Computer Vision and Pattern Recognition}, 2020, pp. 1060--1070.

\bibitem{paschalidou2019superquadrics}
D.~Paschalidou, A.~O. Ulusoy, and A.~Geiger, ``Superquadrics revisited: Learning 3d shape parsing beyond cuboids,'' in \emph{Proceedings of the IEEE/CVF Conference on Computer Vision and Pattern Recognition}, 2019, pp. 10\,344--10\,353.

\bibitem{kawana2020neural}
Y.~Kawana, Y.~Mukuta, and T.~Harada, ``Neural star domain as primitive representation,'' \emph{Advances in Neural Information Processing Systems}, vol.~33, pp. 7875--7886, 2020.

\bibitem{paschalidou2021neural}
D.~Paschalidou, A.~Katharopoulos, A.~Geiger, and S.~Fidler, ``Neural parts: Learning expressive 3d shape abstractions with invertible neural networks,'' in \emph{Proceedings of the IEEE/CVF Conference on Computer Vision and Pattern Recognition}, 2021, pp. 3204--3215.

\bibitem{yang2023topology}
J.~Yang, X.~Jia, and D.-M. Yan, ``Topology guaranteed b-spline surface/surface intersection,'' \emph{ACM Transactions on Graphics (TOG)}, vol.~42, no.~6, pp. 1--16, 2023.

\bibitem{takayama2013sketch}
K.~Takayama, D.~Panozzo, A.~Sorkine-Hornung, and O.~Sorkine-Hornung, ``Sketch-based generation and editing of quad meshes.'' \emph{ACM Trans. Graph.}, vol.~32, no.~4, pp. 97--1, 2013.

\bibitem{marcias2015data}
G.~Marcias, K.~Takayama, N.~Pietroni, D.~Panozzo, O.~Sorkine-Hornung, E.~Puppo, and P.~Cignoni, ``Data-driven interactive quadrangulation,'' \emph{ACM Transactions on Graphics (TOG)}, vol.~34, no.~4, pp. 1--10, 2015.

\bibitem{yu2024super}
J.~Yu and Z.~Wang, ``Super-resolution cloth animation with spatial and temporal coherence,'' \emph{ACM Transactions on Graphics (TOG)}, vol.~43, no.~4, pp. 1--14, 2024.

\bibitem{nuvoli2022skinmixer}
S.~Nuvoli, N.~Pietroni, P.~Cignoni, R.~Scateni, and M.~Tarini, ``Skinmixer: Blending 3d animated models,'' \emph{ACM Transactions on Graphics (TOG)}, vol.~41, no.~6, pp. 1--15, 2022.

\bibitem{de2017regularized}
F.~De~Goes and D.~L. James, ``Regularized kelvinlets: sculpting brushes based on fundamental solutions of elasticity,'' \emph{ACM Transactions on Graphics (TOG)}, vol.~36, no.~4, pp. 1--11, 2017.

\bibitem{knodt2023joint}
J.~Knodt, Z.~Pan, K.~Wu, and X.~Gao, ``Joint uv optimization and texture baking,'' \emph{ACM Transactions on Graphics}, vol.~43, no.~1, pp. 1--20, 2023.

\bibitem{koch2019abc}
S.~Koch, A.~Matveev, Z.~Jiang, F.~Williams, A.~Artemov, E.~Burnaev, M.~Alexa, D.~Zorin, and D.~Panozzo, ``Abc: A big cad model dataset for geometric deep learning,'' in \emph{Proceedings of the IEEE/CVF conference on computer vision and pattern recognition}, 2019, pp. 9601--9611.

\bibitem{wu2021deepcad}
R.~Wu, C.~Xiao, and C.~Zheng, ``Deepcad: A deep generative network for computer-aided design models,'' in \emph{Proceedings of the IEEE/CVF International Conference on Computer Vision}, 2021, pp. 6772--6782.

\bibitem{xu2024cad}
J.~Xu, C.~Wang, Z.~Zhao, W.~Liu, Y.~Ma, and S.~Gao, ``Cad-mllm: Unifying multimodality-conditioned cad generation with mllm,'' \emph{arXiv preprint arXiv:2411.04954}, 2024.

\bibitem{gao2022get3d}
J.~Gao, T.~Shen, Z.~Wang, W.~Chen, K.~Yin, D.~Li, O.~Litany, Z.~Gojcic, and S.~Fidler, ``Get3d: A generative model of high quality 3d textured shapes learned from images,'' \emph{Advances In Neural Information Processing Systems}, vol.~35, pp. 31\,841--31\,854, 2022.

\bibitem{chan2022efficient}
E.~R. Chan, C.~Z. Lin, M.~A. Chan, K.~Nagano, B.~Pan, S.~De~Mello, O.~Gallo, L.~J. Guibas, J.~Tremblay, S.~Khamis \emph{et~al.}, ``Efficient geometry-aware 3d generative adversarial networks,'' in \emph{Proceedings of the IEEE/CVF Conference on Computer Vision and Pattern Recognition}, 2022, pp. 16\,123--16\,133.

\bibitem{schwarz2020graf}
K.~Schwarz, Y.~Liao, M.~Niemeyer, and A.~Geiger, ``Graf: Generative radiance fields for 3d-aware image synthesis,'' \emph{Advances in Neural Information Processing Systems}, vol.~33, pp. 20\,154--20\,166, 2020.

\bibitem{fu2022representing}
T.~Fu, R.~Deng, Y.~Gao, and F.~Zhang, ``Representing scenes as compositional generative neural feature fields based on giraffe for 3d reconstruction of classroom scenes,'' in \emph{International Conference on Intelligent Information Hiding and Multimedia Signal Processing}.\hskip 1em plus 0.5em minus 0.4em\relax Springer, 2022, pp. 227--237.

\bibitem{poole2022dreamfusion}
B.~Poole, A.~Jain, J.~T. Barron, and B.~Mildenhall, ``Dreamfusion: Text-to-3d using 2d diffusion,'' \emph{arXiv preprint arXiv:2209.14988}, 2022.

\bibitem{nichol2022point}
A.~Nichol, H.~Jun, P.~Dhariwal, P.~Mishkin, and M.~Chen, ``Point-e: A system for generating 3d point clouds from complex prompts,'' \emph{arXiv preprint arXiv:2212.08751}, 2022.

\bibitem{shue20233d}
J.~R. Shue, E.~R. Chan, R.~Po, Z.~Ankner, J.~Wu, and G.~Wetzstein, ``3d neural field generation using triplane diffusion,'' in \emph{Proceedings of the IEEE/CVF Conference on Computer Vision and Pattern Recognition}, 2023, pp. 20\,875--20\,886.

\bibitem{wang2023rodin}
T.~Wang, B.~Zhang, T.~Zhang, S.~Gu, J.~Bao, T.~Baltrusaitis, J.~Shen, D.~Chen, F.~Wen, Q.~Chen \emph{et~al.}, ``Rodin: A generative model for sculpting 3d digital avatars using diffusion,'' in \emph{Proceedings of the IEEE/CVF Conference on Computer Vision and Pattern Recognition}, 2023, pp. 4563--4573.

\bibitem{zhao2025hunyuan3d}
Z.~Zhao, Z.~Lai, Q.~Lin, Y.~Zhao, H.~Liu, S.~Yang, Y.~Feng, M.~Yang, S.~Zhang, X.~Yang \emph{et~al.}, ``Hunyuan3d 2.0: Scaling diffusion models for high resolution textured 3d assets generation,'' \emph{arXiv preprint arXiv:2501.12202}, 2025.

\bibitem{xiang2024structured}
J.~Xiang, Z.~Lv, S.~Xu, Y.~Deng, R.~Wang, B.~Zhang, D.~Chen, X.~Tong, and J.~Yang, ``Structured 3d latents for scalable and versatile 3d generation,'' \emph{arXiv preprint arXiv:2412.01506}, 2024.

\bibitem{hyper3d_ai}
{Hyper3D Team}, ``Hyper3d: Ai-powered 3d content generation platform,'' \url{https://hyper3d.ai}, 2025, accessed: 2025-04-06.

\bibitem{kania2020ucsg}
K.~Kania, M.~Zieba, and T.~Kajdanowicz, ``Ucsg-net-unsupervised discovering of constructive solid geometry tree,'' \emph{Advances in Neural Information Processing Systems}, vol.~33, pp. 8776--8786, 2020.

\bibitem{yu2023dualcsg}
F.~Yu, Q.~Chen, M.~Tanveer, A.~M. Amiri, and H.~Zhang, ``Dualcsg: Learning dual csg trees for general and compact cad modeling,'' \emph{arXiv preprint arXiv:2301.11497}, 2023.

\bibitem{zhou2016mesh}
Q.~Zhou, E.~Grinspun, D.~Zorin, and A.~Jacobson, ``Mesh arrangements for solid geometry,'' \emph{ACM Transactions on Graphics (TOG)}, vol.~35, no.~4, pp. 1--15, 2016.

\bibitem{liu2024point2cad}
Y.~Liu, A.~Obukhov, J.~D. Wegner, and K.~Schindler, ``Point2cad: Reverse engineering cad models from 3d point clouds,'' in \emph{Proceedings of the IEEE/CVF Conference on Computer Vision and Pattern Recognition}, 2024, pp. 3763--3772.

\bibitem{blenderkit}
\BIBentryALTinterwordspacing
{BlenderKit}, ``Blenderkit -- 3d asset repository,'' 2025, accessed: 2025-01-20. [Online]. Available: \url{https://www.blenderkit.com/}
\BIBentrySTDinterwordspacing

\bibitem{free3d}
\BIBentryALTinterwordspacing
{Free3D}, ``Free3d -- download free 3d models,'' 2025, accessed: 2025-01-20. [Online]. Available: \url{https://free3d.com/}
\BIBentrySTDinterwordspacing

\bibitem{sketchfab}
\BIBentryALTinterwordspacing
{Sketchfab}, ``Sketchfab -- 3d \& ar visualization platform,'' 2025, accessed: 2025-01-20. [Online]. Available: \url{https://sketchfab.com/}
\BIBentrySTDinterwordspacing

\bibitem{sharma2020parsenet}
G.~Sharma, D.~Liu, S.~Maji, E.~Kalogerakis, S.~Chaudhuri, and R.~M{\v{e}}ch, ``Parsenet: A parametric surface fitting network for 3d point clouds,'' in \emph{European Conference on Computer Vision}.\hskip 1em plus 0.5em minus 0.4em\relax Springer, 2020, pp. 261--276.

\bibitem{chen2024gaussianeditor}
Y.~Chen, Z.~Chen, C.~Zhang, F.~Wang, X.~Yang, Y.~Wang, Z.~Cai, L.~Yang, H.~Liu, and G.~Lin, ``Gaussianeditor: Swift and controllable 3d editing with gaussian splatting,'' in \emph{Proceedings of the IEEE/CVF conference on computer vision and pattern recognition}, 2024, pp. 21\,476--21\,485.

\bibitem{kingma2014adam}
D.~P. Kingma and J.~Ba, ``Adam: A method for stochastic optimization,'' in \emph{ICLR}, 2014.

\end{thebibliography}

\newpage

% \section{Biography Section}
% If you have an EPS/PDF photo (graphicx package needed), extra braces are
%  needed around the contents of the optional argument to biography to prevent
%  the LaTeX parser from getting confused when it sees the complicated
%  $\backslash${\tt{includegraphics}} command within an optional argument. (You can create
%  your own custom macro containing the $\backslash${\tt{includegraphics}} command to make things
%  simpler here.)
 
% \vspace{11pt}

% \bf{If you include a photo:}\vspace{-33pt}
% \begin{IEEEbiography}[{\includegraphics[width=1in,height=1.25in,clip,keepaspectratio]{fig1}}]{Michael Shell}
% Use $\backslash${\tt{begin\{IEEEbiography\}}} and then for the 1st argument use $\backslash${\tt{includegraphics}} to declare and link the author photo.
% Use the author name as the 3rd argument followed by the biography text.
% \end{IEEEbiography}

% \vspace{11pt}

% \bf{If you will not include a photo:}\vspace{-33pt}
% \begin{IEEEbiographynophoto}{John Doe}
% Use $\backslash${\tt{begin\{IEEEbiographynophoto\}}} and the author name as the argument followed by the biography text.
% \end{IEEEbiographynophoto}

\appendices
\section{Elaboration on Operation Parameters}\label{secA1}

\noindent\textbf{Subdivision.} Subdivision level is a non-negative integer ($\mathbb{Z}_{\geq 0}$) indicating the number of surface subdivision iterations. A level of $0$ means no subdivision is applied. 

\noindent\textbf{Extrude.} \emph{Extrude} height is a real number ($\mathbb{R}$) representing the distance that the target faces moves along their average normal. A positive height moves the faces in the direction of the normal, while a negative height moves them in the opposite direction.
\emph{Extrude} width is a positive real number ($\mathbb{R}_{>0}$) acting as a scaling factor along the normal axis. A width of 1 indicates that the extruded face is the same size as the original face; a width less than 1 indicates that the extruded face is smaller than the original face, and vice versa.
Orthogonal angles are two real numbers within $[-\pi, \pi]$, representing rotation angles around two orthonormal basis vectors perpendicular to the normal vector.
Target face index denotes any combination of adjacent faces (or a single face) in the input mesh. Its value range is determined after the approximation stage ends. For example, if the approximated sub-mesh covers faces of indices $\#1$ and $\#3$ that are adjacent to each other, the candidate values are \{(1), (3), (1, 3)\}. Each possible value corresponds to one branch.

\noindent\textbf{Bevel.} Bevel Width is a non-negative real number ($\mathbb{R}_{\geq 0}$) indicating edge smoothing intensity. Segment count is a positive integer representing the number of equal segments for smoothing. Since an excessively large count has a negligible impact on the geometry, we restrict the segment count to a range of 1–9.

\noindent\textbf{Boolean.}  \emph{Boolean} primitive refers to the geometric unit that interacts with the input mesh in \emph{Boolean} operations. Without loss of generality, we determine the types of \emph{Boolean} primitives among cuboid, hexagonal prism, octagonal prism, and cylinder. Primitive affine transformation is represented by a 9-dimensional parameter that includes scale factor ($ \mathbb{R}^3$), translation ($ \mathbb{R}^3$), and rotation angles ($ \mathbb{R}^3$). \emph{Boolean} Type specifies union or difference operations.

\noindent\textbf{Inset.} \emph{Inset} width is a real number in between 0 and 1, representing the ratio of the inset faces size to the original faces size. Since \emph{Inset} is typically combined with \emph{Extrude}, it shares the same input parameter of target face index parameter with \emph{Extrude}, and is excluded from the operation graph during \emph{Extrude} approximation stage.

\noindent\textbf{VertexDisplace.} \emph{VertexDisplace} uses a tensor in $\mathbb{R}^{N \times 3}$ to describe 3D displacement vectors for each vertex in the input mesh, where $N$ is the number of vertices in input mesh.

\noindent\textbf{KnifeCut.} \emph{KnifeCut} operation connects midpoints of a pair of edges on the same face to split the face into two. The indices of edges is specified by the parameter of target edge index. \emph{KnifeCut} is constrained to the initial 10 cycles of the operation graph to prevent mesh wiring from becoming overly complex.

\noindent\textbf{LoopCut.} \emph{LoopCut} operation performs multiple \emph{KnifeCut} along a edge ring. For example, the four vertical edges on the sides of a cube form an edge ring. Performing a \emph{LoopCut} operation on the edge ring would split the four side faces into eight faces. The edge ring is specified by the parameter of target edge ring. \emph{LoopCut} is constrained to the initial 10 cycles of the operation graph to prevent mesh wiring from becoming overly complex.

\noindent\textbf{Mirror.} \emph{Mirror} operation duplicates geometry across a specified axis, which is specified by the parameter of mirror axis (x, y, or z).

\noindent\textbf{SimpleDeform.} \emph{SimpleDeform} includes four sub-operations, each associated with a deformation factor and axis (x, y or z). The factor of \emph{Twist} and \emph{Bend} is an angle in $[-2\pi, 2\pi]$. The factor of \emph{Stretch} and \emph{Taper} is a real number. 

\noindent\textbf{GlobalAffine.} \emph{GlobalAffine} includes three sub-operations including scaling, translation and rotation, each associated with a 3-dimensional parameter ($ \mathbb{R}^3$). \emph{GlobalAffine} is performed on the entire mesh.

\noindent\textbf{EdgeLoopAffine.} \emph{EdgeLoopAffine} applies affine transformations to an edge loop. For example, the four edges of a cube’s top faces form an edge loop. It also includes three sub-operations including scaling, translation and rotation, each associated with a 3-dimensional parameter ($ \mathbb{R}^3$). Note that edge loop and edge ring mentioned in \emph{LoopCut} are distinct elements. They intertwine and collectively form the mesh wiring.

\section{Optimization}
We use Adam \cite{kingma2014adam} for optimization with a learning rate of $0.01$. During optimization, we employ an adaptive progression strategy to govern the depth configuration of the operational graph. The graph is initialized with a minimal base configuration containing $C_0$ cycles of operations ($C_0=7$). Upon detection of optimization stagnation under a moving average criterion over $T$ consecutive iterations, the system automatically appends additional node cycles. This graph expansion mechanism terminates when the reconstruction error converges below a predefined threshold $\epsilon_{\text{term}}$. This strategy enables automatic construction of deep hierarchies for complex geometries while maintaining compact sequence for structurally simple cases. 

To facilitate the progressive graph optimization, we introduce complementary training strategies. 
First, we propose an adaptive visibility modulation scheme, where we associate each point in output results $x\in S_1$ and target $y\in S_2$ with a learnable visibility parameter $z_i\in\mathbb{R}$ to regulate their respective contributions to the loss computation. These parameters are transformed into probabilistic visibility values through temperature-scaled sigmoid function $v_i = \sigma(z_i/\tau_v) \quad (\tau_v=0.3)$. When $v_i \to 0$, the point is ignored. The visibility is incorporated into the loss function by weighting the per-point distance in $\tilde{L}_{S_1}$ (Eqn. 14) and $\tilde{L}_{S_2}$ (Eqn. 15). This strategy allows current nodes to focus on dominant geometric regions while reserving under-optimized regions for subsequent optimization. To prevent degenerate solutions where all $v_i \to 0$, we regulate $v_i$ in the loss function by $\lambda_v\sum{v_i}$.
Second, we impose an additional penalty on the well-aligned region during optimization, such that the well-aligned region remains aligned during the exploration of under-optimized regions. Specifically, for points exhibiting both low distance error and high normal consistency, we amplify their contribution through a weighting multiplier $\omega_{align}>1$ ($\omega_{align}=5$) in the loss computation. This strategy prioritizes optimization of well-aligned regions while maintaining gradient flow through the entire shape.

% \section{Model Training}\label{secA2}

\section{Parallelize Operation Processing}

Our framework is developed in PyTorch. Due to the substantial variations in vertex count and connectivity across different meshes, operations exhibit inherent computational irregularity, impeding uniform tensor manipulation and parallelization. To address this challenge, we extensively refine the implementation of the operations to enable parallel processing of operation branches through dynamic tensor masking, concatenation, and padding techniques. These enhancements enable efficient batched processing of heterogeneous geometric structures, significantly accelerating computational throughput. Empirical results show that our framework attains convergence within approximately 20 minutes for most 3D assets on a single NVIDIA GeForce RTX 3090 GPU.

\section{User study design for 3D modeling professionals}
% questionnaire design → sections → recruitment → validation → ethics
To balance quantitative benchmarks with qualitative insights in evaluating human-AI collaborative tools, we conducted a comprehensive questionnaire-based user study with 3D modeling professionals. The survey instrument is composed of five interconnected modules with 31 closed and open-ended items, assessing the efficacy of 3D-AIGC tool integration within professional workflows.

The questionnaire is structured into the following five modules: background information, needs and pain points, baselines comparison, in-depth method assessment, and subjective feedback.
Module 1 establishes foundational context by gathering professional background information. Participants are asked about their years of experience in 3D modeling, their primary industry focus (\emph{e.g.}, gaming, film, industrial design), and the software tools they use most frequently. These data help segment participants by expertise and workflow preferences, enhancing the credibility of final results.
Module 2 investigates practical challenges and inefficiencies encountered when using current 3D-AIGC tools. Participants report time spent adjusting AI-generated outputs and estimate the overall time saved compared to fully manual modeling. They separately evaluate time savings across three core use cases: static scene object modeling (\emph{e.g.}, tree trunks, mushrooms), character modeling (\emph{e.g.}, animals), and retail product design (\emph{e.g.}, cups, sculptures), reflecting varying workload demands by object type. Participants also rate the frequency of 3D-AIGC usage and identify shortcomings in current tools, such as insufficient fidelity, poor mesh topology, limited editability, or workflow integration issues. Finally, they prioritize which of these areas requires urgent improvement. This module highlights gaps between user expectations and tool performance, providing context for subsequent evaluation.
Module 3 proceeds to a comparative analysis between our method and existing 3D-AIGC methodologies, including: 1) Hunyuan3D-2.0 \cite{zhao2025hunyuan3d}, a state-of-the-art Image/Text-to-3D tool; 2) MeshAnything-V2 \cite{chen2024meshanythingv2}, a high-quality mesh generation method with precise geometric alignment; 3) Point2CAD \cite{liu2024point2cad}, a knowledge-driven CAD sequence reconstruction method; 4) DeepCAD \cite{wu2021deepcad}, a data-driven CAD reconstruction method; 5) GaussionEditor \cite{chen2024gaussianeditor}, a MLLM-based 3D editing method. Participants compare these methods across seven criteria: fidelity (accuracy to input references), mesh wiring (suitability for animation or deformation), step rationality (alignment with human modeling logic), editing flexibility, workflow compatibility, time efficiency and usage intention. This comparison reveals which approaches excel in specific areas and uncovers trade-offs between automation and usability, providing insights into how different technical strategies address real-world design needs.
Module 4 focuses specifically on assessing our proposed method. Participants rate its compliance with critical requirements, including fidelity, mesh wiring, step rationality, editability, and workflow integration. They also estimate the time required to adapt its outputs for static scene object, characters, and retail products, and quantify the time savings compared to manual workflows. This module validates our method’s practicality and benchmarks its performance against the pain points identified earlier, offering a granular view of its strengths and limitations.
Module 5 collects subjective feedback on our method's broader impact. Participants reflect on whether its procedural decomposition aids in understanding model structures, inspires creative brainstorming, or simplifies complex tasks. They also rate their willingness to adopt the algorithm proactively and integrate it into long-term workflows. These qualitative responses gauge user confidence in the tool’s reliability and its potential to enhance, rather than disrupt, existing design processes. Together, the five modules create a holistic evaluation framework, blending quantitative metrics with designer-centric insights to inform the development of 3D-AIGC tools that balance automation, precision, and adaptability.

We recruited professional 3D designers specializing in gaming, film, industrial design, architecture, or digital fabrication through industry forums (Polycount, Blender Artists), professional networks (LinkedIn, ArtStation), and academic collaborations. Within each specialty domain, we oversampled users of Blender and Maya to ensure adequate representation based on industry prevalence. Screening criteria required at least one year of professional 3D modeling experience and recent involvement in relevant 3D projects within the past six months. After applying these filters, 407 qualified designers were enrolled in the user study cohort.

The instrument incorporated established survey methodology practices, featuring cognitive pilot testing with 25 target participants and validation through expert review with 5 senior 3D artists. During pilot testing, selected items were presented as open-ended items without predefined response options (\emph{e.g.}, the first three items in Module 5). Responses from pilot participants were analyzed thematically to inform the systematic conversion into structured Likert scale items. Prior to full user study deployment, validation by 5 senior experts confirmed clarity and logical coherence. To elicit participant preference and mitigate central response bias, Likert scale items used 4-point scales (\emph{e.g.}, fully compliant, mostly compliant, partially compliant, non-compliant) without a neutral midpoint.

\begin{figure*}[!t]
    \includegraphics[width=\linewidth]{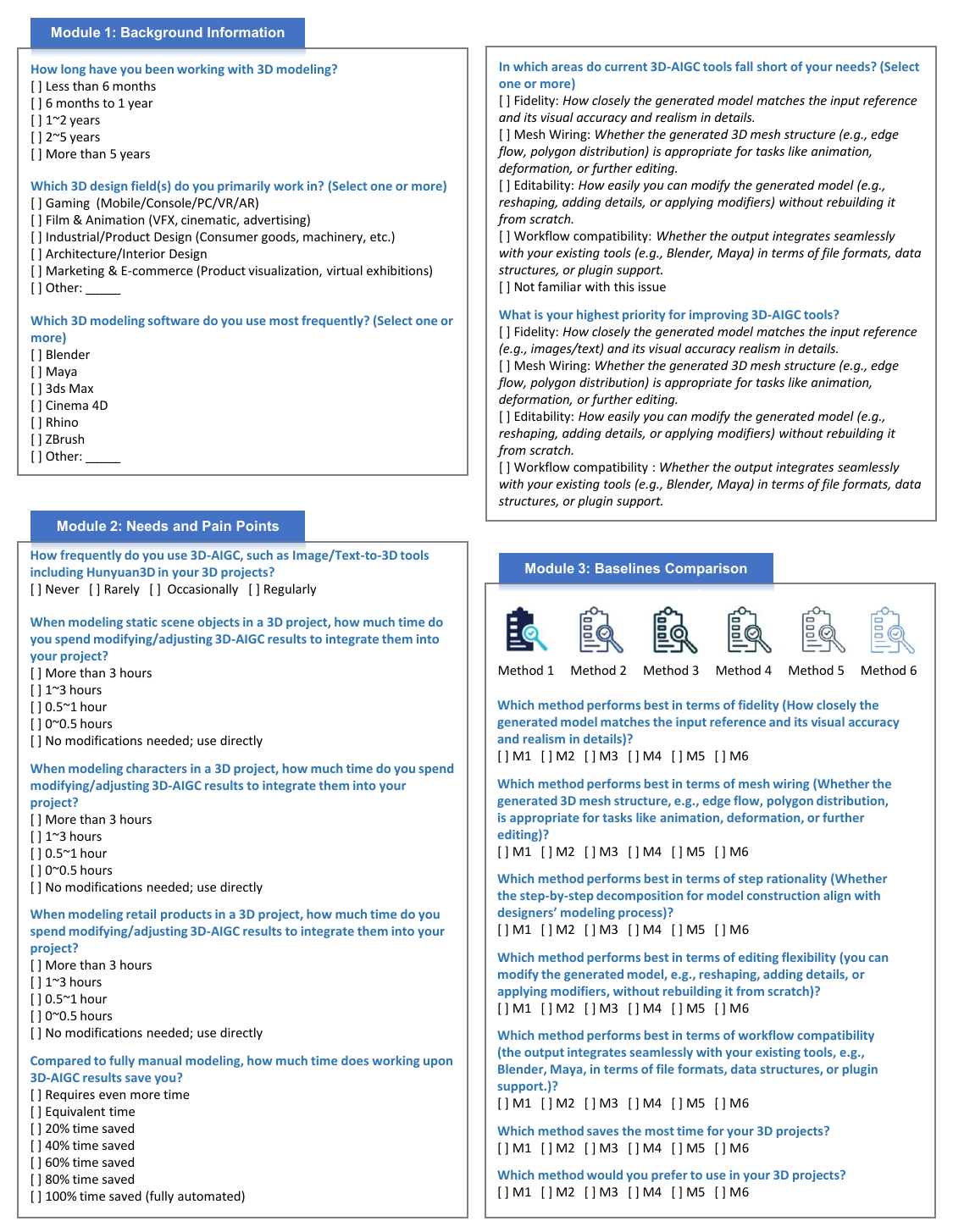}
    \caption{Module 1 - 3 of our user study focus on background information, needs and pain points and baselines comparison.}
    \label{fig:questionnaire_0}
\end{figure*}
\begin{figure*}[!t]
    \includegraphics[width=\linewidth]{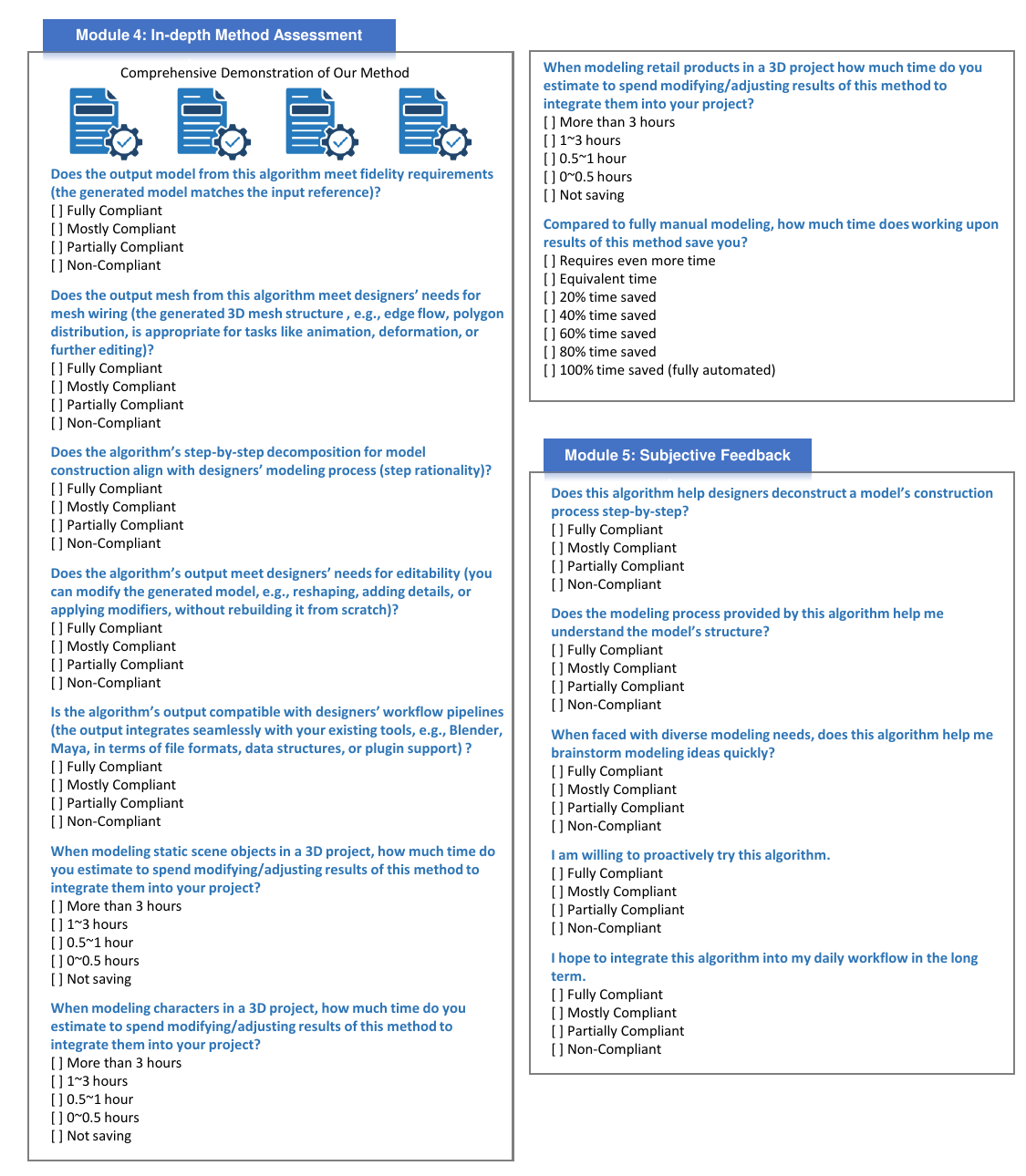}
    \caption{Module 4 - 5 of our user study focus on in-depth assessment of our method and subjective feedback.}
    \label{fig:questionnaire_1}
\end{figure*}

\vfill

\end{document}